\documentclass{aa}  

\usepackage{graphicx}
\usepackage{txfonts}
\usepackage{multirow}
\usepackage[]{hyperref}
\usepackage{natbib}
\usepackage[usenames,dvipsnames]{color}
\usepackage{xspace}
\usepackage[labelfont=bf]{caption}

\newcommand{\Msol}{\ensuremath{\,\mathrm{M_\odot}}\xspace}
\newcommand{\Msolend}{\ensuremath{\,\mathrm{M_\odot}}}

\newcommand{\Rsolend}{\ensuremath{\,\mathrm{R_\odot}}}
\newcommand{\Lsol}{\ensuremath{\,\mathrm{L_\odot}}\xspace}
\newcommand{\Lsolend}{\ensuremath{\,\mathrm{L_\odot}}}
\newcommand{\kk}{\,kK\xspace}

\newcommand{\Myr}{\,Myr\xspace}

\newcommand{\kms}{\,km\,s$^{-1}$\xspace}
\newcommand{\dy}{\,d\xspace}

\newcommand{\GG}[1]{}

\begin{document}

   \title{Wolf-Rayet stars in the Small Magellanic Cloud as testbed for massive star evolution}

   \author{A. Schootemeijer 
            \and
            N. Langer
            }
    \institute{Argelander-Instit\"{u}t f\"{u}r Astronomie, Universit\"{a}t Bonn, Auf dem H\"{u}gel 71, 53121 Bonn, Germany\\ \email{aschoot@astro.uni-bonn.de}
                \\
             }
             
    \authorrunning{Schootemeijer \& Langer}         
   \date{Received September 5, 2017; accepted December 5, 2017}

 
  \abstract
   {The majority of the Wolf-Rayet (WR) stars
represent the stripped cores of evolved massive stars who lost most of their hydrogen envelope.
Wind stripping in single stars is expected to be inefficient in producing WR stars 
in metal-poor environments such as the Small Magellanic Cloud (SMC). 
While binary interaction can also produce WR stars at
low metallicity, it is puzzling that the fraction of WR binaries appears to be about 40\%,
independent of the metallicity.}
   {We aim to use the recently determined physical properties of the twelve known SMC WR stars to explore 
their possible formation channels through comparisons with stellar models.}
   {We used the MESA stellar evolution code to construct two grids of stellar models with SMC metallicity. 
One of these consists of models of rapidly rotating single stars, which evolve in part or completely chemically homogeneously.
In a second grid, we analyzed core helium burning stellar models assuming 
constant hydrogen and helium gradients in their envelopes}
   {We find that chemically homogeneous evolution is not able to account for the majority of the
WR stars in the SMC. However, in particular the apparently single WR star SMC\,AB12, and
the double WR system SMC\,AB5 (HD\,5980) appear consistent with this channel.
We further find a dichotomy in the envelope hydrogen gradients
required to explain the observed temperatures of the SMC WR stars. 
Shallow gradients are found for the WR stars with O\,star companions,
while much steeper hydrogen gradients are required to understand the group of hot apparently single WR stars.}
   {The derived shallow hydrogen gradients in the WR component of the WR+O star binaries are
consistent with predictions from binary models where mass transfer occurs early, in agreement with their
binary properties.
Since the hydrogen profiles in evolutionary models of massive stars become steeper with time after the main sequence, we conclude that most of the hot ($T_\mathrm{eff} > 60$\kk) apparently single WR stars lost their envelope after a phase of strong expansion, e.g., as the result of common envelope evolution with a lower mass companion. 
The so far undetected companions, either main sequence stars or compact objects, are then expected to still be present.
A corresponding search might identify the first immediate double black hole binary progenitor with masses as high as those detected in GW150914.}


   \maketitle
%

\section{Introduction \label{sec:introduction}}
Massive stars can become Wolf-Rayet (WR) stars late in their evolution. These objects are characterized by broad emission lines which originate from a fast, dense stellar wind. WR stars are luminous ($L > 10^{4.5}$\Lsol) and typically very hot and hydrogen depleted, as a result of the removal of a significant part of their hydrogen envelopes. 
With strong stellar winds and dramatic deaths as supernovae they are thought to inject matter processed by nuclear burning into the interstellar medium. Thereby they play an essential role in the chemical evolution of galaxies as well as in providing mechanical and radiative feedback \citep[see e.g.,][]{Hopkins14}. 

Unfortunately, the late phases of massive star evolution are poorly understood, even for stars in our our own galaxy. This is even more so for massive stars in the early universe, which were more metal poor. 
The Small Magellanic Cloud (SMC) is a unique laboratory to study the evolution of low metallicity stars, since its stars 
are metal deficient and as a satellite galaxy of the Milky Way it is sufficiently close for detailed studies of its individual stars. 
Its metal content is around one fifth of the solar value \citep{Venn99, Korn00, Hunter07}, 
which corresponds to that of a spiral galaxies at redshifts $z \approx 3.5$ \citep{Kewley07}.

For lower metallicity, the stellar winds become weaker \citep{Abbott82, Kudritzki87, Mokiem07}. 
Consequently, the winds are less likely to remove the hydrogen envelope,
which raises the question if single stars can become WR stars at all. 
Indeed, it has been proposed that most of the SMC WR stars were formed via envelope stripping by a close binary companion 
\citep{Maeder94, Bartzakos01}. Surprisingly, radial velocity studies  
\citep{Foellmi03, Foellmi04} indicate that the binary fraction of the SMC WR stars is only 40-50$\%$, similar to that in the Milky Way, although this number is based on only twelve sources. 

A possibility to form WR stars from single stars without invoking mass loss is offered by the scenario of 
rotationally induced chemically homogeneous evolution \citep[CHE; see .e.g., ][]{Maeder87, Langer92, Yoon05}.
This channel is indeed expected to work more efficiently for lower metallicity, since then mass loss
induced spin-down, which stops the efficient rotational mixing, is reduced \citep{Langer98}.
CHE has been proposed to lead to long-duration gamma ray bursts \citep{Yoon06,Woosley06}, and, in close binaries, to very massive merging double black holes
\citep{Mandel16, Marchant16} like the gravitational wave source GW150914 \citep{Abbott16}.

Direct empirical evidence for CHE is scarce. 
\cite{Bouret03}, \cite{Walborn04} and \cite{Mokiem06}
find indications for CHE in several very massive O\,stars in the Magellanic Clouds.
\cite{Martins09, Martins13} find CHE to be required to explain the properties of one SMC WR star as well as two WR stars in the Large Magellanic Cloud (LMC) and two WR stars in the Galaxy. \cite{Koenigsberger14}, \cite{Almeida15} and \cite{Shenar17} 
have interpreted observations of different massive close binaries as indications for CHE. 
However, \cite{Hainich15} find that current evolutionary models cannot match all observed properties
of the apparently single WR stars in the SMC. 

To explain the origin of the SMC WR stars is of key importance for the understanding of
massive star evolution at low metallicity. Here, we perform an in-depth theoretical analysis 
of these stars, singling out which of them could result from CHE, and deriving constraints on the
envelope stripping process which might have produced the majority of the remaining WR stars.
This task is greatly facilitated by the recent determination of the stellar parameters of all the 
apparently single \citep{Hainich15} and binary \citep{Shenar16} WR stars in the SMC. 




After providing a brief overview of observational analyses that have been done on the WR stars in the SMC 
up to now in Sect.\,\ref{sec:observations}, we explain the computational method for our analysis in Sect.\,\ref{sec:method}. 
In Sect.\,\ref{sec:rotmix} we show and discuss our results for the rotationally mixed models, and in Sect.\,\ref{sec:other_channels} 
we construct models for stars which have experienced envelope stripping.
We present our conclusions in in Sect.\,\ref{sec:conclusions}.

\section{Empirical properties of Wolf-Rayet stars in the Small Magellanic Cloud \label{sec:observations}}

The first observational overview of WR stars in the SMC, containing four objects, was provided by \cite{Breysacher78}. This number was doubled by \cite{Azzopardi79}, who introduced the nomenclature for the SMC sources with WR characteristics, which we adopt here. The number of known SMC WR stars grew from eight to nine after the work of \cite{Morgan91}. Interestingly, at that time all of the WR stars were thought to have an O-star binary companion due to the presence of hydrogen absorption lines in the spectra. \cite{Conti89} argued however that the presence of these absorption lines could be the consequence of a weaker wind compared to Galactic and LMC WR stars.
After more discoveries \citep{Massey01, Massey03}, radial velocity measurements were performed on all by then twelve SMC WR stars to establish their binary fraction \citep{Foellmi03, Foellmi04}. 
These measurements indicate that only five of the twelve WR stars have a binary companion.


Recently, the stellar parameters of all seven single \citep{Hainich15} and all five binary sources \citep{Shenar16} with a WR star 
in the SMC have been derived using model atmosphere calculations. The derived parameters are listed in 
Table\,\ref{tab:singles} for the single WR stars and in Table\,\ref{tab:binaries} for those in binary systems. 
Due to the, for WR standards, rather weak winds of the SMC WR stars, signified by
the presence of absorption lines in the spectra of most of them, the derived temperatures
and radii are free of the ambiguity which is present in corresponding determinations in more metal-rich
WR stars \citep{Hamann06, Crowther07}.


Fig.\,1 shows the location of the SMC WR stars in the HR diagram. Their luminosities
range from $10^{5.5}\,$L$_{\odot}$ to $10^{6.3}\,$L$_{\odot}$, which implies WR star 
masses of about 15$\dots$60\,M$_{\odot}$ \citep{Langer89}. While the initial mass
range could be identical assuming CHE, their initial masses would have to be roughly in the range
40$\dots$100\,M$_{\odot}$ if they are stripped stars.

Four of the WR stars, that is, the two components of the double WR system SMC\,AB5 (HD\,5980), and the apparently single WR sources SMC\,AB2 and 4 are located to the right of the zero age main sequence
in the HR diagram, while the other nine objects are all considerably hotter. In the following, we will
refer to both groups as to the cool and the hot SMC WR stars, respectively.
Except for SMC\,AB8, which is a WO-type star in a close binary system with a massive O\,star,
all SMC WR stars show significant amounts of hydrogen in their atmosphere. 

In the analysis of the binaries, \cite{Shenar16} found odd properties for SMC\,AB6. In particular,
its luminosity is found to greatly exceed its Eddington luminosity. 
The authors conclude that the observed parameters are probably erroneous due to light contamination by a third star. 
For this reason, we do not consider it later on in our analysis.

\section{Method \label{sec:method}}
We use the detailed one-dimensional stellar evolution code MESA \citep{Paxton11, Paxton13, Paxton15} version 8845 to obtain our stellar models. 

For the initial composition of our SMC models we adopt the one implemented by \cite{Brott11}. Rather than being scaled down uniformly from solar abundances, 
initial abundances of the important elements C, N, O, Mg and Fe are based on different observations in the SMC.
The helium mass fraction of $Y_\mathrm{SMC} = 0.252$ is based on a linear interpolation between the primordial value of $Y = 0.2477$ \citep{Peimbert07} and the solar helium abundance $Y = 0.28$ \citep{Grevesse96} as a function of metallicity.
The opacity tables are obtained from the OPAL opacities \citep{Iglesias96}, using an `effective' metallicity $Z = Z_\odot \cdot (X_\mathrm{Fe, \, SMC} / X_\mathrm{Fe, \, \odot})$. Here, we take the solar values $Z_\odot = 0.017$ and $X_\mathrm{Fe, \, \odot} = 0.00124$ from \cite{Grevesse96} and the $X_\mathrm{Fe, \, SMC}$ value follows from [Fe/H]$_\mathrm{SMC} = -0.6$ from \cite{Venn99}.

The wind mass loss recipe we use also follows \cite{Brott11}, where the choice of prescription depends on the properties of the stellar model. For stars hotter than $\sim$25\kk that have a high surface hydrogen mass fraction of $X_\mathrm{s} > 0.7$, we use the wind recipe of \cite{Vink01}. For hydrogen-poor hot stars with $X_\mathrm{s} < 0.4$, we use the WR mass loss recipe from \cite{Hamann95}, divided by ten to account for wind clumping and downward revisions of the mass loss rate in general (cf. \cite{Yoon05, Yoon06, Brott11}). For stars with in-between $X_\mathrm{s}$ values, $\log \dot{M}$ results from a linear interpolation between both. For all stars cooler than $\sim$25\kk (i.e., the temperature of the bi-stability jump) we use the highest of the values given by the prescriptions from \cite{Vink01} and \cite{Nieuwenhuijzen90}. For all wind prescriptions, we assume a metallicity dependence of $\dot{M} \propto Z^{0.85}$ as in \cite{Vink01}.

\begin{table*}[ht]
\centering    
    \caption[]{Observed parameters of SMC Wolf-Rayet stars in binaries. The values are adopted from \cite{Shenar16}. The orbital period $P_\mathrm{orb}$ and radial velocity amplitudes $K_\mathrm{WR}$ for the WR star and $K_\mathrm{O \, star}$ for the O star (if known) are the values derived by \cite{Foellmi03, Foellmi04}. The exception are the WR stars 5$_\mathrm{A}$ and 5$_\mathrm{B}$ which reside in the same system; their orbital parameters are adopted from \cite{Koenigsberger14}.}
\begin{tabular}{l l l l l l l l l}
\hline
\hline
            \noalign{\smallskip}
            SMC\,AB & $T_*$ & $\log \dot{M}$ & $\log L$ & $X_\mathrm{H}$ & $v_\mathrm{rot}$ & $P_\mathrm{orb}$ & $K_\mathrm{WR}$ & $K_\mathrm{O \, star}$ \\
             & [kK] & [\Msol yr$^{-1}$] & [\Lsol] & & [km s$^{-1}$] & [d] & [km s$^{-1}$] & [km s$^{-1}$] \\
            \noalign{\smallskip}
            \hline
            \noalign{\smallskip}
            3 & $78^{+5}_{-5}$ & $-5.3^{+0.1}_{-0.1}$ & $5.93^{+0.05}_{-0.05}$ & $0.25^{+0.05}_{-0.05}$ & - &10.1 & 144 & -\\
            \noalign{\smallskip}
            5$_\mathrm{A}$ & $45^{+5}_{-5}$ & $-4.5^{+0.1}_{-0.1}$ & $6.35^{+0.10}_{-0.10}$ & $0.25^{+0.05}_{-0.05}$ & $<300$ & \multirow{2}{*}{19.3} & 214 & - \\
            \noalign{\smallskip}
            5$_\mathrm{B}$ & $45^{+10}_{-7}$ & $-4.5^{+0.3}_{-0.3}$ & $6.25^{+0.15}_{-0.15}$ & $0.25^{+0.20}_{-0.20}$ & $<400$ & & 200 & - \\
            \noalign{\smallskip}
            6 & $80^{+15}_{-10}$ & $-5.1^{+0.2}_{-0.2}$ & $6.28^{+0.10}_{-0.10}$ & $0.4^{+0.1}_{-0.1}$ & - & 6.5 & 290 & 66\\
            \noalign{\smallskip}
            7 & $105^{+20}_{-10}$ & $-5.0^{+0.2}_{-0.2}$ & $6.10^{+0.10}_{-0.10}$ & $0.15^{+0.05}_{-0.05}$ & - & 19.6 & 196 & 101\\
            \noalign{\smallskip}
            8 & $141^{+60}_{-20}$ & $-4.8^{+0.1}_{-0.1}$ & $6.15^{+0.10}_{-0.10}$ & $0.0^{+0.15}$ & - & 16.6 & 176 & 55\\
            \noalign{\smallskip}
            \hline
            \label{tab:binaries}

\end{tabular}
\end{table*}

\begin{table}
\centering    
    \caption[]{Observed parameters of apparently single SMC Wolf-Rayet stars. All values are adopted from \cite{Hainich15}. 
    }
\begin{tabular}{l l l l l l}
\hline
\hline
            \noalign{\smallskip}
            SMC\,AB & $T_*$ & $\log \dot{M}$ & $\log L$ & $X_\mathrm{H}$ & $v_\mathrm{rot}$  \\
             & [kK] & [\Msol yr$^{-1}$] & [\Lsol] & & [km s$^{-1}$] \\
            \noalign{\smallskip}
            \hline
            \noalign{\smallskip}
            1 & $79^{+6}_{-6}$ & $-5.58^{+0.2}_{-0.2}$ & $6.07^{+0.2}_{-0.2}$ & $0.5^{+0.05}_{-0.05}$  & $< 100$\\
            \noalign{\smallskip}
            2 & $47^{+3}_{-3}$ & $-5.75^{+0.2}_{-0.2}$ & $5.57^{+0.1}_{-0.2}$ & $0.55^{+0.05}_{-0.05}$ & $< 50$\\
            \noalign{\smallskip}
            4 & $45^{+3}_{-3}$ & $-5.18^{+0.2}_{-0.2}$ & $5.78^{+0.1}_{-0.2}$ & $0.25^{+0.05}_{-0.05}$ & $< 100$\\
            \noalign{\smallskip}
            9 & $100^{+6}_{-6}$ & $-5.65^{+0.2}_{-0.2}$ & $6.05^{+0.2}_{-0.2}$ & $0.35^{+0.05}_{-0.05}$ & $< 200$\\
            \noalign{\smallskip}
            10 & $100^{+6}_{-6}$ & $-5.64^{+0.2}_{-0.2}$ & $5.65^{+0.2}_{-0.2}$ & $0.35^{+0.05}_{-0.05}$ & $< 200$\\
            \noalign{\smallskip}
            11 & $89^{+6}_{-6}$ & $-5.56^{+0.2}_{-0.2}$ & $5.85^{+0.2}_{-0.2}$ & $0.4^{+0.05}_{-0.05}$ & $< 200$\\
            \noalign{\smallskip}
            12 & $112^{+6}_{-6}$ & $-5.79^{+0.2}_{-0.2}$ & $5.90^{+0.2}_{-0.2}$ & $0.2^{+0.05}_{-0.05}$ & $< 200$\\
            \noalign{\smallskip}
            \hline
            \label{tab:singles}

\end{tabular}
\end{table}

In convective zones, mixing is modeled according to the standard mixing-length theory \citep{Bohm58}. We use a mixing-length parameter $\alpha_\mathrm{MLT} = 1.5$. 
The convective boundaries are set by the Ledoux criterion for convection. Convective overshooting above the convective core is treated with a step overshoot parameter. We adopt $\alpha_\mathrm{ov} = 0.335$, as calibrated with the rotational velocities versus $\log g$ \citep{Brott11} of a large sample of LMC stars observed with the VLT-FLAMES survey \citep{Evans05}. In the layers that are stable to convection according to the Ledoux criterion but not according to the Schwarzschild criterion, we assume that semiconvection takes place with an efficiency of $\alpha_\mathrm{sc} = 1$ \citep{Langer91}.

Rotationally enhanced mass loss is implemented as a function of the ratio of the stellar rotation to the critical rotation velocity \citep{Friend86}: the $\dot{M}$ boost factor is set to $( 1 / (1 - w))^\xi$, where $w = \varv / \varv_\mathrm{crit}$ and $\xi = 0.43$. For the efficiency of rotational mixing we use $f_\mathrm{c} = 1/30$, which is in agreement with calibrations of \cite{Brott11} to nitrogen enrichment in rotating stars analyzed by \cite{Hunter08}.

In their analysis of SMC WR stars, \cite{Hainich15} and \cite{Shenar16} provided a temperature $T_*$ which is defined in a fashion similar to the effective temperature: at a radius $R_*$, defined as the radius where the Rosseland optical depth $\tau = 20$, $T_*$ satisfies the equation $T_* = (L / (4 \pi \sigma R_*^2 ))^{1/4}$. Here, $L$ is the luminosity of the star and $\sigma$ is Boltzmann's constant.

Therefore, in our models we also calculate $T_*$ at 
$\tau = 20$, taking wind optical depth into account.
The latter is calculated using Eq. (11) in \cite{Langer89}. This formula assumes electron scattering opacity, but the effect on the resulting $T_*$ is negligible for our WR stars with SMC metallicity. We note that the difference between this $T_*$ and the effective temperature $T_\mathrm{eff}$ is typically smaller than a few percent in our models.



   \begin{figure}
   \centering
   \includegraphics[width = \linewidth]{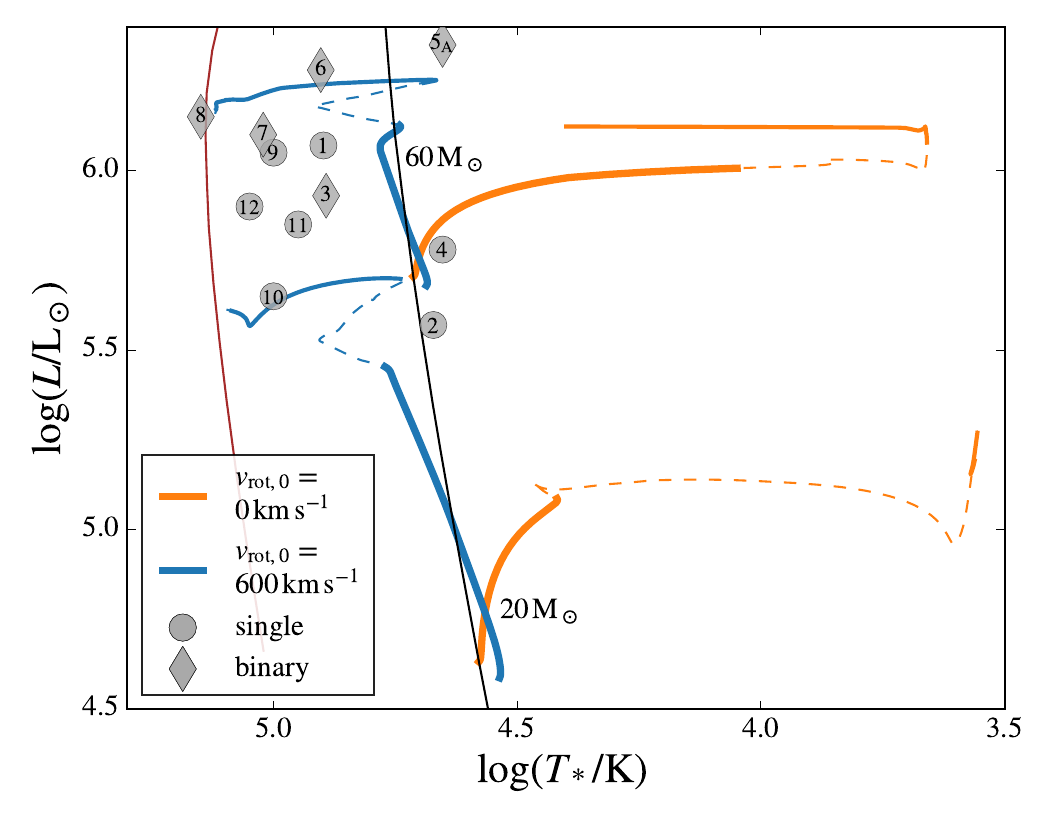}
   \caption{Hertzsprung-Russell diagram with tracks of MESA models with initial masses of 20\Msol and 60\Msol and different initial rotation velocities. 
   The black line represents the zero-age main sequence for stars with the composition described in Sect. \ref{sec:method}, while the brown line represents the zero-age main sequence for helium stars.
   Thick solid lines indicate that a model is core hydrogen burning with $X_\mathrm{c} > 0.01$; thin solid lines indicate core helium burning; dashed lines indicate that a model is in an in-between, shorter-lived phase. 
   The observed apparently single Wolf-Rayet stars (Table\,\ref{tab:singles}) are displayed as gray circles. Those in a binary system are displayed as gray diamonds (Table\,\ref{tab:binaries}).
   The numbers indicate the identifier of the star, e.g., SMC\,AB1.}
             \label{fig:some_tracks}%
    \end{figure}

\section{Rotationally mixed models \label{sec:rotmix}}
To demonstrate the effect that rapid rotation has on our massive star models, we show two distinct sets of tracks in Fig.\,\ref{fig:some_tracks}. The evolutionary tracks are shown for models which have no rotation and models which have a high initial rotation velocity of 600\kms. The fast-rotating models are able to avoid the significant expansion of the hydrogen envelope, as they are evolving chemically (quasi-)homogeneously. In this section, we compare the observed SMC WR stars to models that are in the core hydrogen burning phase (Sect.\,\ref{sec:coreh}) and the core helium burning phase (Sect.\,\ref{sec:corehe}). The reason we focus on these two phases is that the chance that a significant fraction of the SMC WR stars is in any other phase is small: both phases combined make up over 99$\%$ of the total stellar lifetime. In Appendix D  we provide an overview of the best fits to the observed stars for both families of models.

We explore the mass range $M_\mathrm{ini} = 20, \, \ldots , 100$\Msol with 5\Msol intervals (10\Msol intervals above 70\Msol). 
The initial rotation velocities of the models cover the range $\varv_\mathrm{rot, ini} = 350, \, \ldots , 600$\kms with 10\kms intervals.


   \begin{figure}
   \centering
   \includegraphics[width = \linewidth]{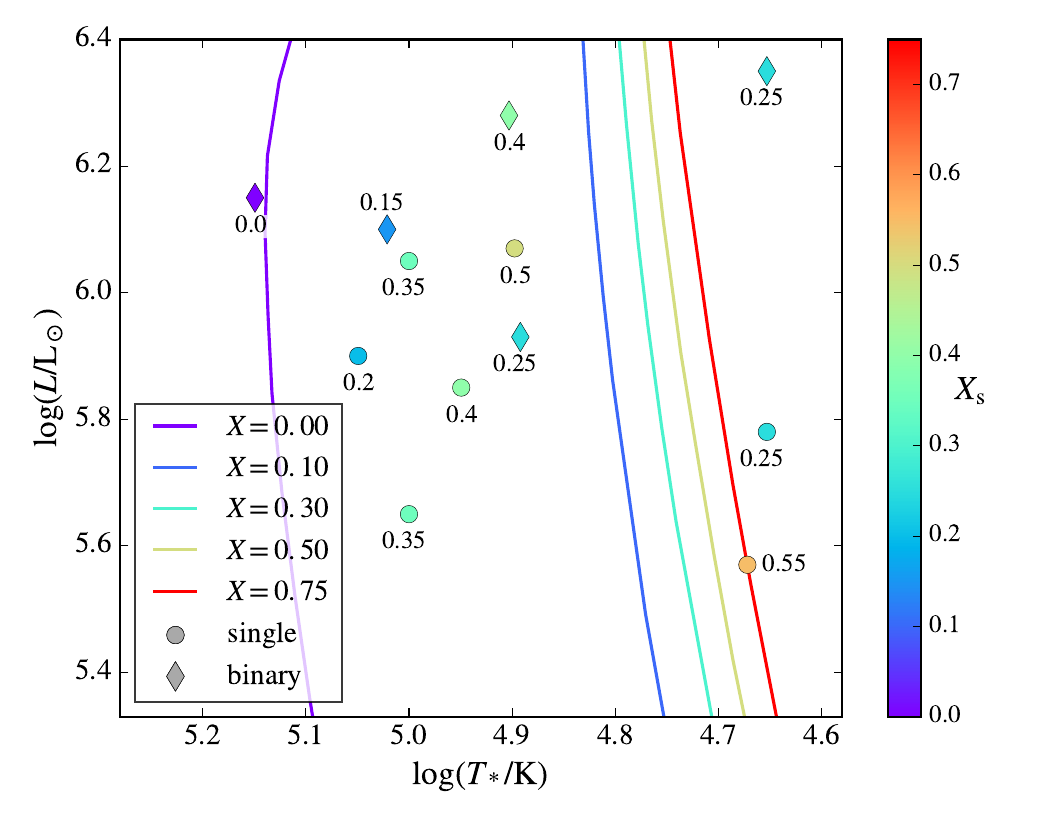}
   \caption{Hertzsprung-Russell diagram with lines indicating the positions of homogeneous stellar models with hydrogen mass fractions $X$ of $0, \, 0.1, \, 0.3, \, 0.5,$ and $0.75$. The metallicity $Z$ is as described in Sect. \ref{sec:method}, while the helium mass fraction $Y$ is given by $Y = 1 - X - Z$. The models with $X = 0$ are helium burning, the others are hydrogen burning. The numbers near the scatter points indicate the surface hydrogen mass fractions $X_\mathrm{s}$ of the observed SMC WR stars. Both the lines and the scatter points are color coded for $X_\mathrm{s}$. Circles indicate apparently single stars; diamonds indicate binaries.}
              \label{fig:zamss}%
    \end{figure}

\subsection{Core hydrogen burning phase \label{sec:coreh}}
As is shown in Fig.\,\ref{fig:some_tracks}, the core hydrogen burning models do not reach the high temperatures that are observed for nine out of twelve SMC WR stars. The same tendency emerges in Fig.\,\ref{fig:zamss}, where chemically homogeneous SMC models with different hydrogen mass fractions are displayed. This figure implies that even hydrogen-poor chemically homogeneous stars are cooler than these nine hot SMC WR stars. 

Evolutionary models of rotationally mixed stars are not completely chemically homogeneous because the mixing is not infinitely fast. However, our models that experience blueward evolution always have a surface and central hydrogen abundance with a difference of $X_\mathrm{s} - X_\mathrm{c} < 0.1$. Therefore, the homogeneous models shown in Fig.\,\ref{fig:zamss} have a chemical profile comparable to these rotationally mixed models.

When comparing the observed stars to chemically homogeneous models with the same surface hydrogen mass fraction $X_\mathrm{s}$, the observed stars can be as much as 0.3 dex hotter (i.e., $100$\kk vs $\sim$50\kk for SMC\,AB 10). The hydrogen-free models in Fig.\,\ref{fig:zamss} are considerably hotter than models which contain hydrogen, as they have contracted until temperatures high enough for helium ignition were reached.

Apart from the temperatures, there is a modest conflict between the observed upper limits on the rotation velocities of the hot apparently single SMC WR stars and the rotational velocities of the models. Although depending on initial rotation velocity and angular momentum loss, the models typically retain $\varv_\mathrm{rot} > 250$\kms; the upper limits on $v \sin i$ of these stars are 100-200\kms.

The terminal-age main sequence (TAMS), that is, the point where hydrogen is exhausted in the core, is followed by
a short contraction phase in which the models do reach higher temperatures (Fig.\,\ref{fig:some_tracks}). However, this phase is short lived ($\tau \approx \tau_\mathrm{MS}/1000$) and during the contraction the star spins up to even higher rotation velocities. As a result, the likelihood that the observed hot SMC WR stars are contracting stars that have just evolved past the main sequence is very small.

The objects that are not too hot to be core hydrogen burning are the apparently single stars SMC\,AB2 and 4 as well as both WR stars in the binary system SMC\,AB5. For the two single stars, the rotation velocities are with $\varv\sin i < 50$\kms (AB2) and $\varv\sin i < 100$\kms (AB4) relatively well constrained. Although the models spin down during their evolution, we find that it is unlikely that the low observed rotation velocities of the stars are an inclination effect. The models for which we achieve a best fit using the observed parameters $T_*$, $L$ and $X_\mathrm{s}$ have rotation velocities of 302 and 183\kms for SMC\,AB2 and 4, respectively. Then, following the formula provided by \cite{Grin17} we calculate that the chance that the observed $v\sin i$ limit is not exceeded is 1.4$\%$ for SMC\,AB2 and 16$\%$ for SMC\,AB4. The probability to observe both stars pole-on enough at the same time therefore seems marginal.


   \begin{figure}
   \centering
   \includegraphics[width = \linewidth]{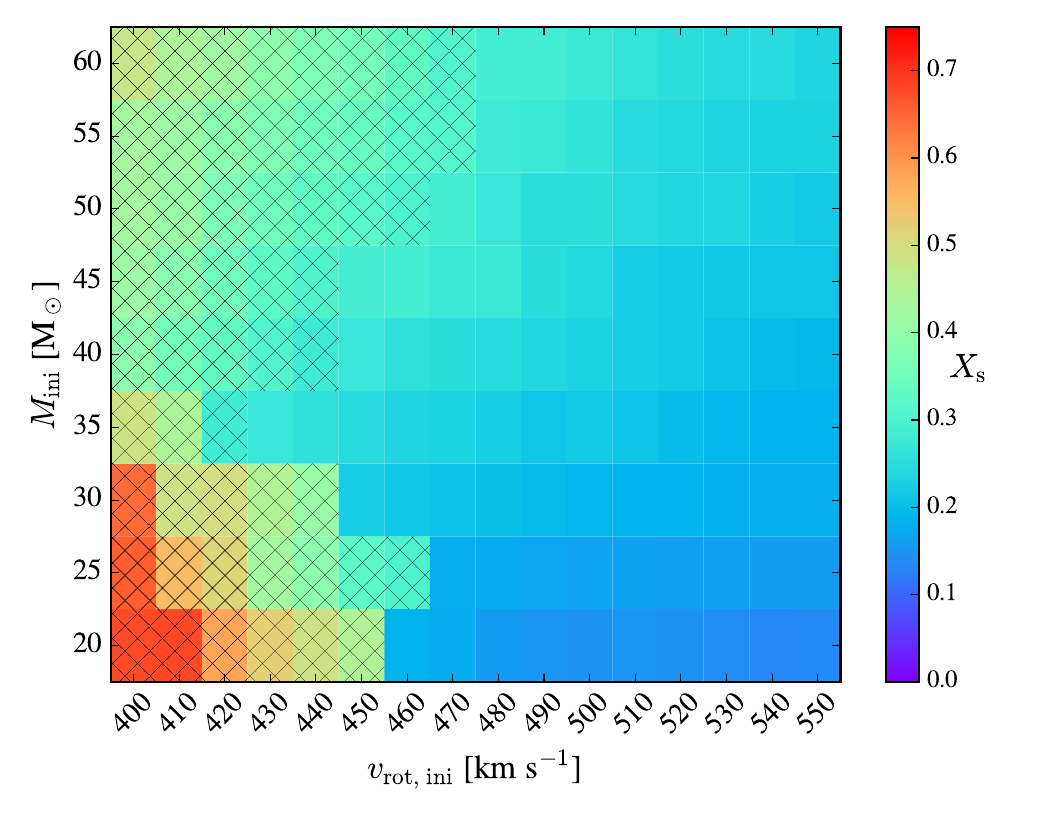}
   \caption{Diagram showing which models can become hot during helium burning. Each rectangle represents a model sequence with a certain initial rotation velocity and mass. The color indicates the surface hydrogen mass fraction $X_\mathrm{s}$ at the moment helium is ignited in the core. For the hatched systems, $T_*$ does not overlap with the observed temperature range of the SMC WR stars during the entire core helium burning phase.}
              \label{fig:xs_ighe}%
    \end{figure}


Moreover, \cite{Vink17} have used spectropolarimetry to search for hints of rapid rotation in SMC WR stars. These were only found in the double WR system SMC\,AB5, which has high upper limits on $\varv\sin i$ (Table\,\ref{tab:binaries}). Thus, their findings are in agreement with the upper limits from \cite{Shenar16}.

Both WR stars in the source SMC\,AB5 are slightly on the cool side of the homogeneous models in Fig.\,\ref{fig:zamss}. 
The observed parameters can be reproduced more accurately (i.e., all within $1\sigma$, Table\,\ref{tab:chi2_bin_h}) by models with intermediately rapid rotation, in which rotational mixing becomes inefficient as they spin down during their evolution (e.g., the $\varv_\mathrm{rot, \, 0} = 400$\kms track in Fig.\,\ref{fig:hrd_vroti}). Alternatively, the temperatures can be lower than expected due to envelope inflation that can occur in very luminous stars \citep{Sanyal17}. It is worth mentioning that the effective temperature of SMC\,AB5$_\mathrm{A}$ was unstable in the recent past: after a luminous blue variable type eruption in 1994, it has increased from $\sim$25\kk to its current value of $\sim$45\kk \citep{Georgiev11}. The high observed upper limits on $\varv \sin i$ are not in conflict with the rotational velocities of the models. Therefore, we conclude that these WR stars are in agreement with core hydrogen burning stars going through CHE, as was proposed by \cite{Koenigsberger14}. 

\subsection{Core helium burning phase \label{sec:corehe}}
The surface hydrogen mass fraction $X_\mathrm{s}$ at the moment core helium burning commences, which we show in Fig.\,\ref{fig:xs_ighe}, depends on the initial mass and rotation velocity of the stellar model. As could be expected, it shows that initially more rapidly rotating models have lower $X_\mathrm{s}$ values. 

Because rotational mixing is not infinitely efficient, the hydrogen envelope will have a shallow abundance gradient (Sect.\,\ref{sec:coreh}) and therefore also a gradient in the mean molecular weight $\mu$.
In addition, due to stellar spin-down as a result of angular momentum loss via stellar winds, the mixing can become inefficient enough for the $\mu$ gradient to build up, which further inhibits mixing. This way, models which initially evolve almost homogeneously are able to retain intermediate surface hydrogen mass fractions through their core helium burning phase. 
For less massive stars, on the one hand a higher initial rotation velocity is required to mix the stellar interior to the surface. On the other hand, they have weaker stellar winds which result in less spin-down. As a result of these effects, the window for intermediate $X_\mathrm{s}$ values during core helium burning narrows down with lower masses. 

\begin{figure}
   \centering
   \includegraphics[width = \linewidth]{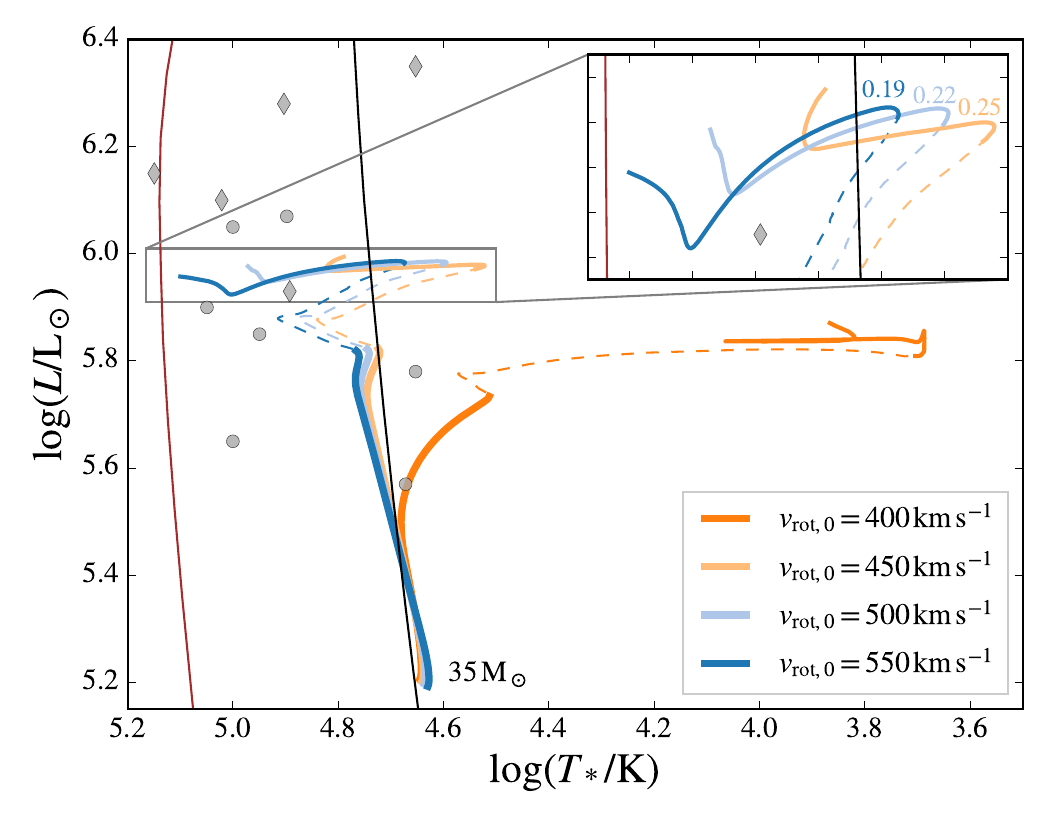}
   \caption{Same as Fig.\,\ref{fig:some_tracks}, but now 35\Msol models with various initial rotation velocities are displayed. The inset at the top right corner zooms in on the core helium burning phase. The numbers near the tracks in the inset indicate their surface hydrogen mass fraction $X_\mathrm{s}$ at the moment helium is ignited in the core.}
              \label{fig:hrd_vroti}%
    \end{figure}

When CHE is discontinued before the final stages of the main sequence evolution, a significant amount of hydrogen is retained and the star is unable to avoid the giant phase.
This scenario is exemplified in Fig.\,\ref{fig:hrd_vroti} by the stellar model with $\varv_\mathrm{rot, \, 0} = 400$\kms. We find that models which have $X_\mathrm{s} \geq 0.3$ are cooler than the observed SMC WR star with the lowest $T_*$ at all times during core helium burning. This means that the group of hydrogen-rich hot single WR stars (SMC\,AB 1, 9, 10, 11) and the binary WR star SMC\,AB6 do not match helium burning models which went through CHE, since all have $X_\mathrm{s} \geq 0.35$. 


The low temperature of these helium burning models with $X_\mathrm{s} \geq 0.3$ is related to their hydrogen profile. 
Because this hydrogen profile has a shallow gradient in the rotationally mixed models, the hydrogen envelope extends deep into the star - as a result, the star has a large radius. We discuss hydrogen profiles in more detail in Sect.\,\ref{sec:other_channels}.

In contrast to their hydrogen-rich counterparts,
core helium burning models with $X_\mathrm{s} \lesssim 0.25$ are able to reach the $T_*$ that is observed for the cool WR stars. This is the case for the $\varv_\mathrm{rot, \, 0} = 450$\kms model in Fig.\,\ref{fig:hrd_vroti}. However, it is not able to reach a value of $T_*$ as high as observed for the hot stars. Models with higher initial rotation velocities are able to do so, but they have lower surface hydrogen mass fractions on the order of $X_\mathrm{s} \approx 0.2$. Therefore, core helium burning quasi-CHE models are able to explain all observed properties of the relatively hydrogen-poor SMC WR stars. These include the hot single WR star SMC\,AB12 ($X_\mathrm{s} = 0.2$) and those in WR+O binaries SMC\,AB3, 7 and 8 ($X_\mathrm{s} = 0.25, \, 0.15\, \mathrm{and} \, 0.0$ respectively). For these objects, we are able to find solutions where the models meet the observed parameters $T_*$, $L$ and $X_\mathrm{s}$ simultaneously within $1\sigma$. 

After their late core hydrogen burning phase, our models spin down enough for the upper limits on $\varv \sin i$ to agree with $\varv_\mathrm{rot}$ of the core helium burning models. A downside of this core helium burning quasi-CHE scenario however is that this phase is relatively short-lived: $\sim$5$\%$ of the core hydrogen burning timescale. This would imply that for every core helium burning object, $\sim$20 less evolved core hydrogen burning stars would be present in the population which are going through the same evolutionary scenario. Although these could be missed in observational campaigns due to a variety of biases (e.g., lower luminosity during core hydrogen burning, detectability of helium enrichment), this poses a potential problem. Previous observations of O-type and early B-type stars in the SMC \citep{Mokiem06, Penny09, Bouret13} indicate that 
their rotational velocity distribution is skewed to higher values than the rotational velocity distribution of their Galactic counterparts. However, the difference is modest.
More extended surveys would be required to resolve this question in the future.

\section{Stripped stars \label{sec:other_channels}}
As a next step we investigate whether helium burning stars that are partly stripped of their hydrogen envelopes can account for the observed properties of the SMC WR stars. The hydrogen envelope is defined as `all layers that still contain hydrogen'. In principle, stripping of the envelope can be done by stellar winds, by short-lived outbursts of strong mass loss, or by a binary companion. We do not model the complete evolution of stars for these scenarios; instead we investigate a grid of stellar models with a helium core and a variety of hydrogen envelopes to compare with the observed parameters of SMC WR stars. 

\begin{figure}
   \centering
   \includegraphics[width = \linewidth]{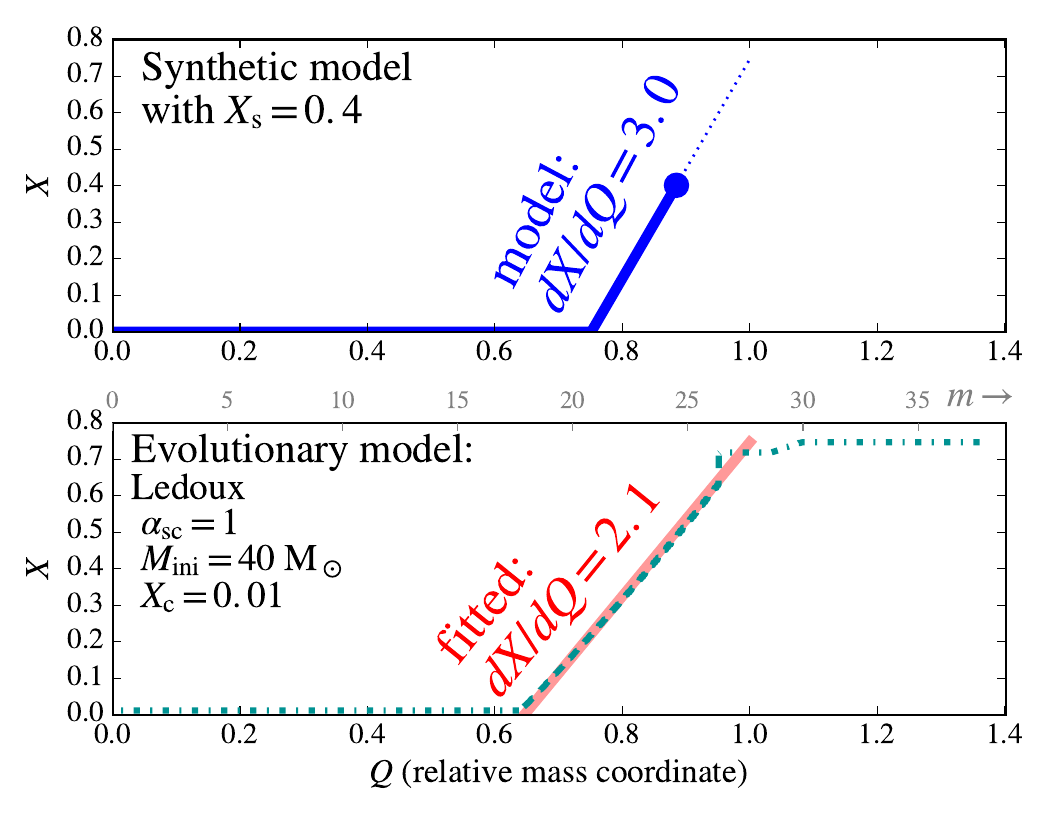}
   \caption{\textbf{Top:} Example of a synthetic hydrogen profile (solid line). The blue dot represents the surface. This model has a slope of $dX/dQ = 3.0$ and surface hydrogen fraction $X_\mathrm{s} = 0.4$. \textbf{Bottom:} Hydrogen profile of a 40\Msol star model near the end of core hydrogen burning. Plotted in red is the fitted slope of $dX/dQ = 2.1$. The part of the profile that was used for fitting is displayed with a dashed line, the rest is displayed with a dash-dotted line. For comparison, the mass coordinate $m$ (in units of \Msol) is also displayed at the top of this panel.
   \label{fig:schematic_dxdq}}%
    \end{figure}

To characterize the hydrogen envelopes of these models we use two parameters: the hydrogen mass fraction at the stellar surface $X_\mathrm{s}$ and the slope of the hydrogen profile $dX/dQ$. To illustrate our method, we show an example of such a synthetic hydrogen profile in Fig.\,\ref{fig:schematic_dxdq} (top). Here, $Q$ is a normalized mass coordinate, with $Q=0$ in the stellar center, and with $Q=1$ defined as the mass coordinate 
where the linear slope of the hydrogen profile reaches $ X(Q) = X_0$ (the hydrogen mass fraction at the zero-age main sequence; $X_\mathrm{0, \, SMC} = 0.746$). At the surface (blue dot) this model has a hydrogen fraction of $X_\mathrm{s} = 0.4$. 
A model in which a layer with $ X_\mathrm{s} < X_0$ is exposed necessarily has a $Q$ value at its surface of $Q_\mathrm{s} < 1$. The value for $Q_\mathrm{s}$ is given by
\begin{equation}
    Q_\mathrm{s} = 1 - \frac{ X_\mathrm{0} - X_\mathrm{s} }{dX/dQ}.
\end{equation}
Then, the hydrogen mass fraction $X(Q)$ throughout a star which is hydrogen-depleted at the surface in 
the range $0 \leq Q \leq Q_\mathrm{s}$ is:
\begin{equation}
    X(Q) = \max \Big[ 0, \; X_\mathrm{s} - \big( Q_\mathrm{s} - Q \big) \, dX/dQ \Big].
    \label{eq:x}
\end{equation}

At first glance the definition of the variable $Q$ might seem overly complicated. 
However, it avoids that the value of the hydrogen profile slope becomes dependent on the stellar mass 
or on the surface hydrogen mass fraction $X_\mathrm{s}$.


For comparison, we show a hydrogen profile of a 40\Msol evolutionary model at the end of hydrogen core burning in the bottom of Fig.\,\ref{fig:schematic_dxdq}. 
As the mass of the convective core in massive main sequence stars is decreasing rather linearly as function of time, it leaves a hydrogen profile $X(Q)$ with a rather constant hydrogen gradient. 
Therefore, we are able to fit a slope with $dX/dQ = 2.1$ that closely represents the hydrogen profile in the model. 
We fit the slope of the hydrogen profile until a `plateau' with a constant hydrogen fraction is encountered which contains more than 5\% of the mass of the star.
The part of the model that we use for the fit 
in the bottom of Fig.\,\ref{fig:schematic_dxdq}
thus has a hydrogen fraction $0.01 < X < 0.7$. 
$Q=1$ is set by the point where the fitted slope $dX/dQ = 2.1$ reaches $X_\mathrm{0}$.
At core hydrogen exhaustion, $Q = 1$ typically coincides with the border of the convective core at the zero-age main sequence.

While hydrogen profiles in evolutionary models may be more complex than what we assume for our synthetic models, our approach is the first order approximation
and contains the minimum number of parameters. Furthermore, it does not assume any evolutionary history and may thus cover 
scenarios that are not usually dealt with, like common envelope evolution or a stellar merger.

The stellar models computed for this section are simulated with the same physics assumptions as described in 
Sect.\,\ref{sec:method}, but with different initial chemical profiles. The hydrogen mass fraction $X(Q)$ inside 
the star follows from the surface hydrogen fraction $X_\mathrm{s}$ and the adopted $dX/dQ$ value as described 
by Eq.(\,\ref{eq:x}). The metallicity $Z$ also corresponds to the value declared in Sect.\,\ref{sec:method}, 
but since all material should have been processed by nuclear burning we assume CNO equilibrium. Finally, 
the helium mass fraction $Y$ follows from $Y = 1 - X - Z$. The models are evolved until they 
have a central helium mass fraction of $Y_\mathrm{c} = 0.75$, while mixing and abundance changes due 
to nuclear burning in the hydrogen envelope are switched off.
We note that in stellar envelopes as hot as those of the SMC WR stars, little mixing is expected to occur.

\subsection{Inferred hydrogen profiles in SMC WR stars \label{sec:h_profiles}}

In this section, we compare stellar models with various hydrogen profile slopes to the observed single and binary SMC WR stars. 
On average, the apparently single SMC WR stars are more hydrogen rich than their binary counterparts by $\Delta X_\mathrm{s} > 0.1$ (Tables  \ref{tab:binaries} \& \ref{tab:singles}). 
Naively one might expect that these more hydrogen-rich stars are also cooler, since hydrogen-free stars move toward the helium main sequence. 
Surprisingly, the observed temperatures show the opposite trend.


\begin{figure}
\centering
\includegraphics[width = \linewidth]{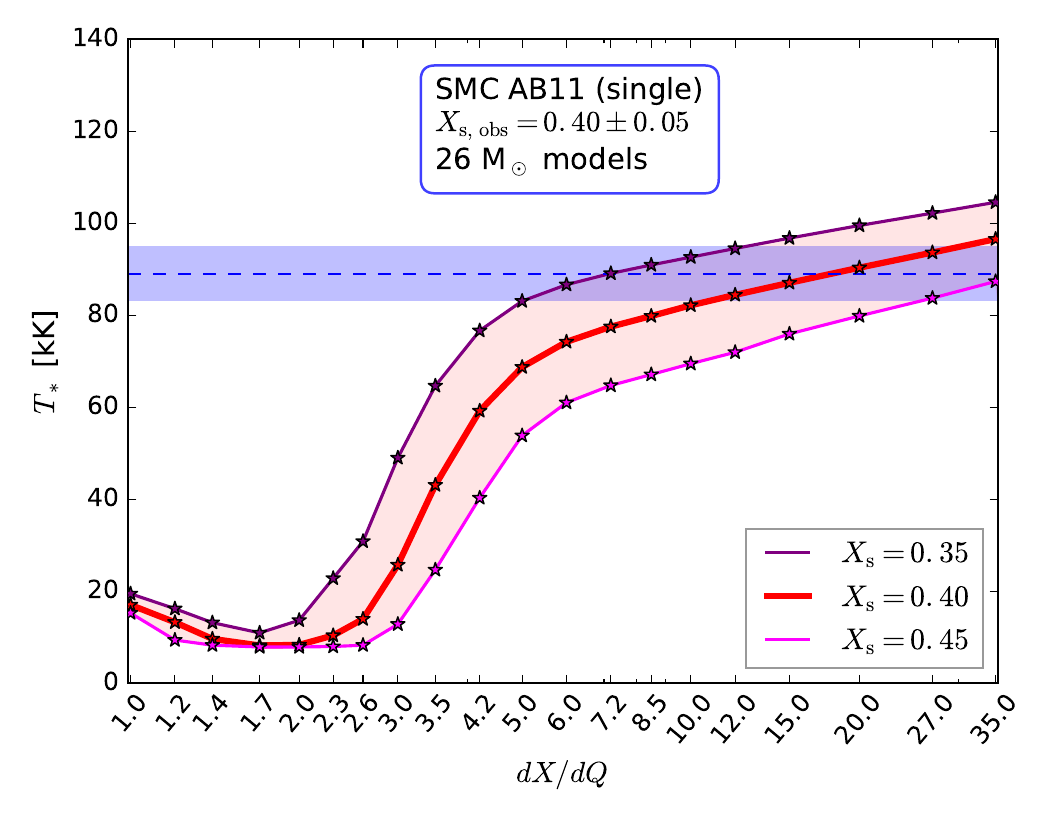}
\caption{Diagram showing the best-fitting hydrogen slope $dX/dQ$ for SMC\,AB11. The temperature $T_*$ is plotted as a function of $dX/dQ$ for models with surface hydrogen mass fractions $X_\mathrm{s}$ which correspond to $X_\mathrm{s, \, obs} \pm 1 \sigma$. Each `star' symbol represents a stellar model. 
The blue dashed line represents observed $T_*$ of SMC\,AB 11 to which the models are compared, whereas the blue shaded area represents the error margin. We note that the x-axis is in log scale. \label{fig:dxdq_teff11}}%
\end{figure}

Our method is exemplified by Fig.\,\ref{fig:dxdq_teff11}, where we compare our models with the apparently single star SMC\,AB11. Each model in this figure has a helium core and a hydrogen profile which follows from the 
adopted slope $dX/dQ$ and the surface hydrogen abundance $X_\mathrm{s}$, which is chosen close to 
the observed one, 
$X_\mathrm{s, \, obs}$. 
The mass of the models is chosen such that their luminosity matches the observed luminosity. 
We consider hydrogen profile slopes in the range of $1 < dX/dQ < 35$. 

Fig.\,\ref{fig:dxdq_teff11} demonstrates that models with shallow slopes ($dX/dQ < 4$), which contain more hydrogen, 
are far cooler than the observed WR star. 
The $dX/dQ$ range in which models with a surface hydrogen mass fraction of $X_\mathrm{s, \, obs} \pm 1\sigma$ 
are able to reproduce the observed temperature $T_\mathrm{*, \, obs} = 89$\kk covers $7 < dX/dQ < 35$ with a best-fitting value of $dX/dQ \simeq 20$. 
These values are much higher than in the example model we showed in Fig.\,\ref{fig:schematic_dxdq}, where we found $ dX/dQ = 2.1$ for a star at the end of core hydrogen burning.

We repeat this exercise for the other SMC WR stars. For these objects, similar plots are provided in Appendix B,
and the results of the fits are provided in Table\,\ref{tab:dxdq_fit}. We display the $dX/dQ$ range of models 
that match the observed surface temperatures of the WR stars in Fig.\,\ref{fig:dxdq_teff_fits}. 
They are divided into `hot' objects (their $T_*$ only matches helium burning models) and `cool' objects 
(which can be both, helium and hydrogen burning). Below, we discuss the results for the  apparently single WR stars (Fig.\,\ref{fig:dxdq_teff_fits}, top) and for the WR stars in binaries (Fig.\,\ref{fig:dxdq_teff_fits}, bottom). 

\begin{itemize}
    \item \textbf{Apparently single stars:} although for the five hot apparently single WR stars the scatter 
around the best-fitting $dX/dQ$ value is rather large, we conclude that values on the order of $dX/dQ \approx 2$ 
can be ruled out according to our models (except for SMC\,AB12, which is marginally consistent with a shallow slope). 
However, all of them are in agreement with $dX/dQ$ values of $10-15$ or even larger. SMC\,AB1 is most extreme, as it 
needs $dX/dQ$ values of $\sim$35 or more to approach $T_\mathrm{*,  obs}$. 
SMC\,AB12 could also be explained via CHE (cf., Sect.\,4).


   \begin{figure}
   \centering
   \includegraphics[width =\linewidth]{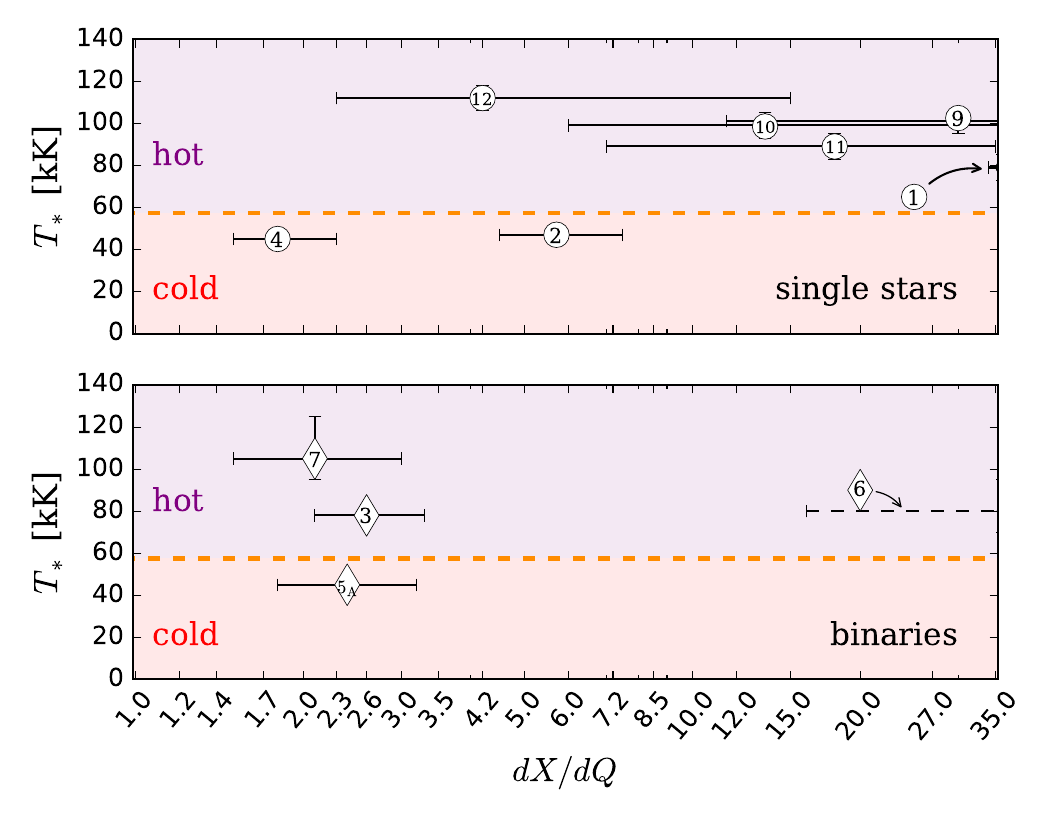}
   \caption{Best-fitting $dX/dQ$ values for apparently single SMC WR stars (top) and those in binaries (bottom). SMC\,AB9 \& 10 have the same $T_\mathrm{*, \, obs}$, but they are slightly offset for clarity. The numbers indicate the identifiers of the stars. The peculiar source SMC\,AB6 is displayed with a dashed line.
   \label{fig:dxdq_teff_fits}}%
    \end{figure}

The two remaining apparently single stars are the two cool single stars. These are SMC\,AB2 and 4, which are the two only apparently single SMC WR stars for which we find that they also have a temperature compatible with core hydrogen burning. The $T_*$ of SMC\,AB2 matches the observed temperature only for a narrow interval around $dX/dQ \approx 6$; 
SMC\,AB4 matches to helium burning stripped star models.

    \item \textbf{Binaries:} here, it seems to be the other way around: all binary WR stars 
    fit to models with shallow slopes, with the best fits occurring around $2 < dX/dQ < 3$. Therefore, unlike for the apparently single stars, this indicates that their hydrogen profile slopes are similar to those of the TAMS star model shown in Fig.\,\ref{fig:schematic_dxdq}. 
    We again mention that for SMC\,AB5, the observed parameters could also be explained by both components going through CHE (cf., Sect.\,4). Finally, the temperature of SMC\,AB8 is in agreement with that of a helium star. However, since no hydrogen has been detected at its surface we cannot consider the hydrogen profile of this star.
\end{itemize}

In summary, we find that the observed parameters of the SMC binary stars are  similar to those of a 40\Msol model stripped at the TAMS. On the other hand, we infer much steeper hydrogen profiles for the group of hot apparently single SMC WR stars. We also visualize this in Appendix A, where we provide HRDs showing models with the best-fitting $dX/dQ$ values to the observed binary and hot apparently single SMC WR stars. In Table\,\ref{tab:dxdq_fit} we show the lifetimes of the inferred hydrogen envelopes, which we find to be in the same order as the core helium burning timescales.

\subsection{Progenitor evolution and binary status of the SMC WR stars \label{sec:evolution}}

To put the hydrogen slopes derived in the last section into perspective, we now investigate $dX/dQ$ values of evolutionary models.
To do this comprehensively is beyond the scope of this paper, since it relates to uncertain internal mixing process and their
triggering, for example, by rotation or binary interaction \citep[cf.][]{Langer12}. 
For this purpose, we only consider single star models. That is, we 
do not model how these stars lose their hydrogen envelopes, but we 
discuss what the products of envelope stripping would look like.

\subsubsection{Terminal-age main sequence} \label{sec:tams}
We compute nonrotating single star models in the initial mass range $15\Msol\dots 110\Msol$
up to core hydrogen exhaustion, with the same assumptions as the models discussed in Sect.\,4. 
These models cover the observed luminosity range of the SMC WR stars.
Apart from models with our default value for the overshooting parameter $\alpha_\mathrm{ov}$, 
we also compute models for extreme cases, that is, 
with $\alpha_\mathrm{ov} = 0.15$ and $\alpha_\mathrm{ov} = 0.5$.

\begin{table}
\centering    
    \caption[]{Values for the fitted slope of the hydrogen profile $dX/dQ$ for models with different initial masses and different values of the overshooting parameter $\alpha_\mathrm{ov}$. All models are at the end of core hydrogen burning.}
\begin{tabular}{l | l l l l l l l}
\hline
\hline
            \hfill $M_0$ [M$_\odot$] & 15 & 25 & 40 & 60 & 85 & 110 & avg.\\
            \hline
            $\alpha_\mathrm{ov} = 0.15$ & 1.5 & 1.8 & 1.8 & 1.8 & 1.7 & 1.7 & 1.7\\
            $\alpha_\mathrm{ov} = 0.335$ (default) & 1.8 & 2.0 & 2.1 & 2.0 & 2.1 & 2.1 & 2.0\\
            $\alpha_\mathrm{ov} = 0.5$ & 2.1 & 2.3 & 2.4 & 2.4 & 2.4 & 2.4 & 2.3\\
            \hline
\end{tabular}
\label{tab:dxdq}
\end{table}

Until the end of hydrogen burning, the slope of the hydrogen profile in the models is relatively constant, due to the constant recession speed of the convective core. 
Table\,\ref{tab:dxdq} shows the values of $dX/dQ$ derived for models with different initial parameters,
at the time of core hydrogen exhaustion. Noticeably, the variation of $dX/dQ$ with initial mass
is very small. We find that the core overshooting parameter has a larger impact. 
However, overall the TAMS 
hydrogen profile slopes are restricted to the narrow range $1.5 \leq dX/dQ \leq 2.4$. 

We consider this result to be robust. It is likely not sensitive to the adopted criterion for convection
or to the efficiency of semiconvective mixing \citep[][see also Fig.\,\ref{fig:logr_dxdq}]{Langer85}. 

Also the mass loss rates are so small that their uncertainty is not relevant here. However, in rapid rotators, the hydrogen gradient may be significantly different (cf., Sect.\,4).

\subsubsection{Post-main-sequence evolution \label{sec:pms}}

The post-main-sequence evolution of single stars involves more uncertainties
than their main sequence evolution, many of which also affect the hydrogen profile.
Here, we cannot systematically explore the whole parameter space, but restrict ourselves to
elaborate on one emerging trend: namely that, in most considered cases, the hydrogen profile becomes steeper with time.
We find this to be the case due to several effects.

After core hydrogen exhaustion the star contracts as a whole. As a consequence, a hydrogen burning shell source
is ignited, which drives the expansion of the hydrogen-rich envelope. During this stage,
convective and semiconvective regions form, at first above the hydrogen burning shell.
Later on, when the star becomes a cool supergiant, envelope convection can occur and extend
down into the region of varying hydrogen concentration, also known as dredge-up.
Since all these mixing processes push hydrogen into deeper layers, 
that is, closer to hydrogen-depleted layers, the hydrogen profile becomes steeper as a consequence.
The efficiency of this mixing, which is controlled by semiconvection, 
is poorly known, and as a consequence also the post-main-sequence radius evolution of massive stars is uncertain
(e.g., \cite{Langer85}).


   \begin{figure}
   \centering
   \includegraphics[width =\linewidth]{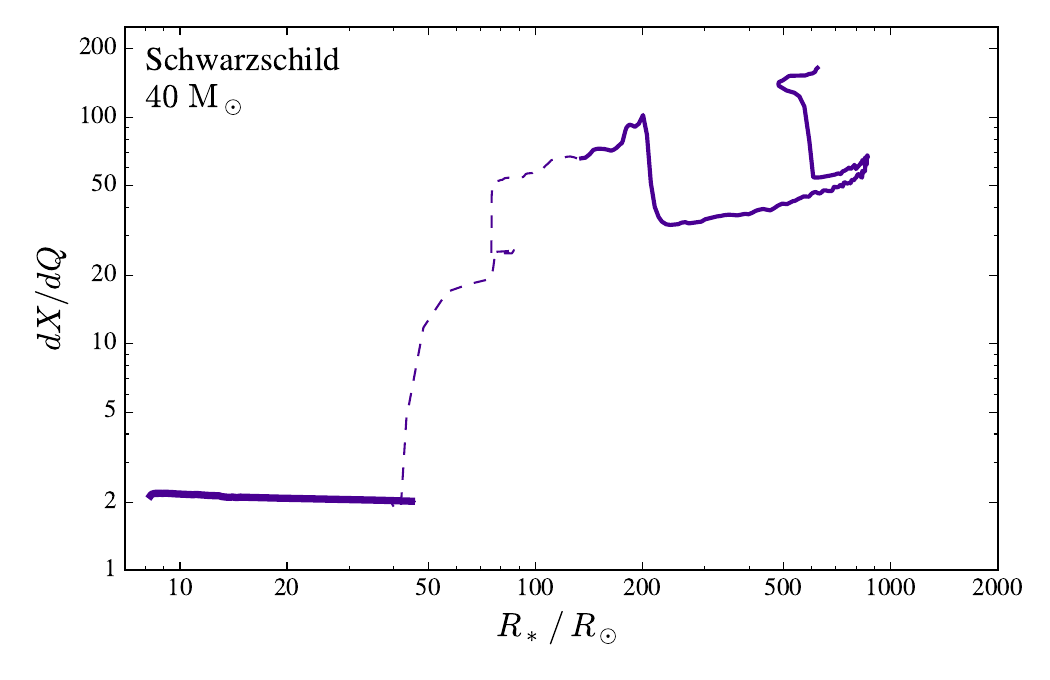}
   \caption{Evolution of the hydrogen slope $dX/dQ$ of a 40\Msol model as a function of stellar radius. As in Fig.\,\ref{fig:some_tracks} and Fig.\,\ref{fig:hrd_vroti}, the hydrogen burning phase is shown with a thick solid line, the short-lived hydrogen shell burning phase with a dashed line and helium burning with a thin solid line. This sequence was computed adopting the Schwarzschild criterion for convection. \label{fig:logr_dxdq}}
    \end{figure}

Yet, an increase of the steepness of the hydrogen profile after the ignition of the hydrogen burning shell is expected in any case. We illustrate this in Fig.\,\ref{fig:logr_dxdq}, where we show the $dX/dQ$ value as function of the stellar radius during the core helium burning evolution of a 40\Msol stellar model, computed with the Schwarzschild criterion for convection. While the evolution starts with the TAMS value of $dX/dQ\simeq 2$, values of the order of 20 to 50 are achieved during the blue supergiant stage, while maximum values near 50 or even larger are obtained in the red supergiant stage. 

\begin{figure}
   \centering
   \includegraphics[width =\linewidth]{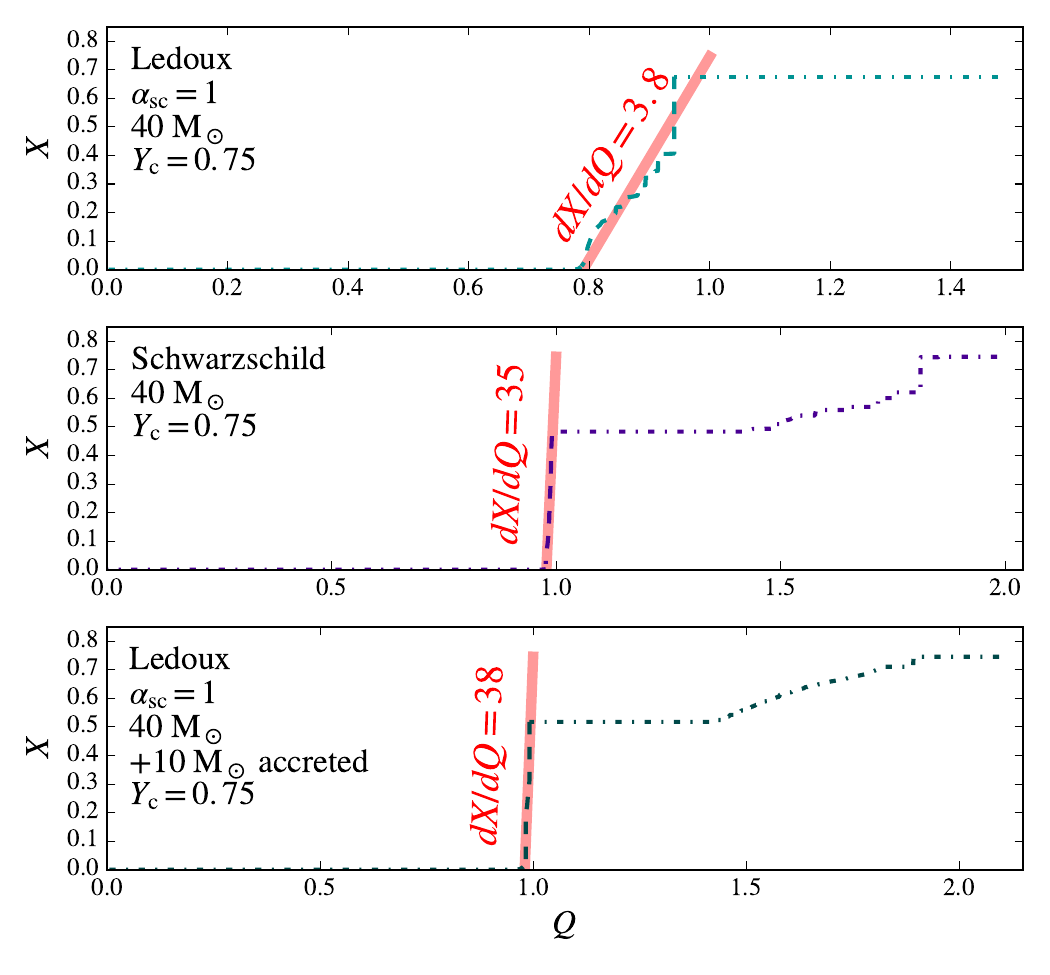}
   \caption{Same as Fig.\,\ref{fig:schematic_dxdq}, but showing core helium burning models with central helium mass fraction of $Y_\mathrm{c} = 0.75$. Apart from the model with default assumptions (top), we also show a model where the Schwarzschild criterion for convection is assumed (middle) and a model which has accreted 10\Msol near the end of the main sequence (bottom). \label{fig:quadruplot}}
    \end{figure}

A hydrogen profile of a model from this sequence where core helium burning has advanced to a core helium mass fraction of $Y_\mathrm{c} = 0.75$ is shown in the middle panel of Fig.\,\ref{fig:quadruplot}. It can be compared to a corresponding plot for a model computed with the Ledoux criterion for convection and a semiconvection parameter of $\alpha_{\rm sc}=1$, where the mixing is much more limited, 
and $dX/dQ$ does increase only to values of about five (top panel of Fig.\,\ref{fig:quadruplot}).




We compute the evolution of a 40\Msol model which accreted 10\Msol of unprocessed material near 
the end of its main sequence evolution, which may simulate the situation of a mass gainer in a close binary
system \citep{Braun95}, using again $\alpha_{\rm sc}=1$. 
The corresponding hydrogen profile is shown in the bottom panel of Fig.\,\ref{fig:quadruplot}.  
As a result of the mass accretion, its convective core has 
grown in mass, which also gives rise to a very steep hydrogen profile.

Another process which steepens the hydrogen profile is hydrogen shell burning. However, this does not cause changes in $dX/dQ$ which are as dramatic as those caused by the mixing processes discussed above. In the most extreme case this could cause an increase of $dX/dQ = 2$ to $dX/dQ = 4$, but the effect vanishes for large $dX/dQ$ values (see Appendix C).

\subsection{Connecting the hydrogen profile in SMC WR stars with their evolutionary history}

WR stars, in particular the ones in the SMC discussed here, have lost most of their hydrogen-rich envelope.
The high temperature of the hot majority of the SMC WR stars requires them to be in the stage of core
helium burning. In the following, we discuss the various possibilities for how these stars could have lost
their hydrogen-rich envelope, and, according to our discussion above, which slope for the hydrogen gradient we
could expect in the corresponding scenario.

\subsubsection{Single star mass loss} 
As shown above, the small $dX/dQ$ values of the WR stars in binaries imply that their 
progenitors have been stripped in a compact stage (see also Sect.\,\ref{sec:rlo}).
The much higher $dX/dQ$ values of the apparently single WR stars indicate that their progenitors strongly expanded before their envelopes were stripped (Sect.\,\ref{sec:pms}).
Since envelope stripping is expected to produce WR stars also in wide binaries \citep{Schneider15} --- in about as many cases as in Case\,A or early Case\,B, given the \cite{Sana12} orbital period distribution --- we conclude that at least the majority of the apparently single WR stars lost their envelope due to a binary companion (cf. Sect\,\ref{sec:ce}.)

We may still ask the question whether a fraction of these stars might have lost their envelopes
as single stars, that is, without the help of a companion.
At present, they (apart from SMC\,AB4) 
are losing about $2$\Msolend\Myr$^{-1}$ through stellar winds. At this rate, which is higher 
than what is expected for the main sequence evolution, mass loss is not strong enough to 
expose hydrogen-depleted layers for reasonable initial masses. As stellar wind mass loss rates during the hot stages of evolution may currently even be overestimated at low metallicity \citep{Hainich17}, hot star winds can not have removed the hydrogen envelopes of the SMC WR stars.

Thus, the envelope stripping in any single star scenario would have to occur in the 
cool part of the HR diagram, possibly in the form of episodic luminous blue variable (LBV)-type mass loss,
or in a yellow/red supergiant stage. Since the Small Magellanic Cloud does not host red supergiants 
with a luminosity as high as those of the bulk of the
SMC WR stars \citep{Blaha89, Neugent10, Yang12}
the loss of the hydrogen-rich envelope during a RSG phase is therefore unlikely.
The LBV phenomenon, on the other hand, has been associated with the stellar Eddington limit \citep[e.g.,][]{Lamers88, Sanyal15}, which, at SMC metallicity, is located well above $10^6\,$L$_{\odot}$ 
(\cite{Ulmer98} --- see also \cite{Sanyal17}, their Fig.\,2, for the inflation of TAMS stars which are at the Eddington limit).
Therefore, in particular the apparently single SMC WR stars can not be expected to have hit the Eddington limit.
However, as long as the LBV mass loss is not fully understood --- for example, it has been proposed recently that the LBV phenomenon is caused by binary interaction \citep{Smith15} --- a single star origin of {\em some} of the apparently single SMC WR stars can not be fully excluded.

Independent of the mass loss mechanism, also the luminosity distribution of the SMC WR stars 
appears not easily compatible with any single star WR formation channel. If such a channel exists,
it would work above a certain mass, or luminosity, threshold. As the binary channels work
for all masses, we would expect both contributions above the threshold luminosity, but only
WR binaries below it. However, observations show that the five most luminous SMC WR stars are all in binaries 
and the five least luminous ones are all apparently single (Table\,\ref{tab:binaries}\&\ref{tab:singles}).
We conclude that a single star origin of the apparently single SMC WR stars can not be
firmly excluded, but does appear unlikely.

\subsubsection{Stable Roche lobe overflow} \label{sec:rlo}
When the initially more massive star in a close binary expands and fills its Roche lobe,
mass transfer commences. For relatively small initial orbital periods (Case\,A and early
Case\,B mass transfer), the mass transfer is stable in the sense that contact is avoided
\citep{Podsiadlowski92, Wellstein01},
and only stops after most of the hydrogen-rich envelope has been stripped off
the mass donor. Since stable mass transfer also requires initial mass ratios close to one,
the outcome would be a stripped star, that is, a nitrogen sqeuence WR (WN) star in close orbit with 
an O\,star of nearly equal or higher mass. 

The WN binaries SMC\,AB3 and AB7 can be well explained
in this way, and the WO+O binary SMC\,AB8 fits the same scenario, 
only that the system is further evolved \citep[in agreement with][]{Shenar16}.
Their short present-day orbital periods (in the range of 6-20\dy, Table\,\ref{tab:binaries}), indicate that they experienced binary interaction during or shortly after their main sequence evolution.

Interestingly, the WR stars in the WN+O binaries have, according to our analysis above, 
shallow hydrogen slopes with $dX/dQ\simeq 2$ (Fig.\,\ref{fig:dxdq_teff_fits}). Since in mass donors of
Case\,A and early Case\,B binaries the mixing processes which can increase the slope of
the hydrogen profile (see above) have not yet occurred by the time of mass transfer
\citep[see for example Fig.\,2 of][]{Wellstein99}, such shallow slopes are indeed
expected for stable mass transfer systems. Such evolution appears therefore most likely
for the WN+O binaries in the SMC. 

We note that the binary SMC\,AB5 (HD\,5980), which consists of
two very hydrogen-deficient stars, cannot be explained well by stable mass transfer.
As discussed by \cite{Koenigsberger14}, chemically homogeneous evolution, perhaps tidally
induced \citep{deMink09, Marchant16}, can explain this binary best.

\subsubsection{Common envelope evolution} \label{sec:ce}
Mass transfer in initially relatively wide binaries (late Case\,B, Case\,C), and/or with
initial mass ratios very different from one does lead to contact and the formation of a
common envelope (CE). The outcome of this will be a merger in many cases, which may lead
back to the single star scenario discussed above. However, when the more evolved star
has significantly expanded before the CE evolution sets in, the hydrogen-rich envelope
may be bound loosely enough such that an envelope ejection occurs and the binary as such
survives \citep{Ivanova13}.
The result will be a WN star with, most likely, a main sequence companion. Here, 
if the primary could expand sufficiently, the 
mass of the main sequence companion may even be rather small \citep{Kruckow16}.

This scenario could apply to the hot apparently single WR stars, that is,
SMC\,AB1, and SMC\,AB9...12, since a low or intermediate mass companion might have been easily missed
in the binary search of \cite{Foellmi03}. 
Furthermore, the large $dX/dQ$ values
which we found for all these stars imply that the WR progenitors did expand to large radii
before they were stripped. 
In view of Fig.\,\ref{fig:logr_dxdq} we may even speculate that intermediate values of $dX/dQ$
are rare because they would correspond to binaries of intermediate initial periods,
where a merger is the most likely outcome.

%

\subsubsection{Reverse mass transfer} 
In some binaries, a WR star may also form from the initially less massive star. For this
to happen, the binary must have survived an earlier phase of, most probably, stable mass transfer
(see above), and the initially more massive star (e.g., the WR star in the WR+O binaries discussed above)
will likely be a compact object by the time reverse mass transfer starts. Due to the large mass ratio
in this situation, the system will likely undergo CE evolution at this stage, out of which it
may emerge as a WR binary with a compact companion (WR+cc). 

Again, this scenario may apply to the apparently single WR stars, because for not too large compact object masses
radial velocity variations may remain small. While the compact companion may accrete matter
from the wind of the WR star and become a strong X-ray source, the X-ray emission may also be
weak, in particular if the compact object is a black hole and when the formation of an accretion disk is avoided
\citep{Shapiro76}. 
Since the secondary star has accreted
substantial amounts during the first, stable, mass transfer phase, the resulting WR star can be
expected to show a steep hydrogen profile (cf.\,Fig.\,\ref{fig:quadruplot}, bottom).

Thus the hot apparently single WN stars in the SMC could also
be well explained through this scenario. Since some of the WR+O binaries will merge or might break up
when the first supernova occurs, the number of the WR+cc binaries is expected to be smaller
than that of WR+O binaries. Nevertheless, the reverse mass transfer scenario offers the chance
that among the apparently single WR stars in the SMC, we may soon detect the first direct double black hole progenitor with black holes as massive as those recently detected by LIGO \citep{Abbott16, Abbott16b, Abbott17}.

\subsection{The only hydrogen-free SMC WR star - SMC AB8}
The fact that the only hydrogen-free star SMC WR star (SMC\,AB8) is of the type WO (i.e., it has also lost its pure helium envelope) could be seen as an indication for efficient semiconvection. As a result of semiconvective mixing, the CO core is able to grow and will be buried less deep in the He mantle, making it easier to expose. Given its current mass loss rate it would lose $\sim$ 6\Msol during core helium burning. We find that for a scenario with inefficient semiconvection, only helium stars with $M < 15-20$\Msol develop sufficiently low mass helium envelopes above the CO core. This mass is well below the mass expected for its luminosity, which is around 40\Msol \citep{Grafener11}.

\section{Conclusions \label{sec:conclusions}}
In this study, we have compared the observed parameters of Wolf-Rayet (WR) stars in the Small Magellanic Cloud (SMC) 
with a variety of models obtained with the detailed stellar evolution code MESA. 
This includes evolutionary models which undergo chemically (quasi-)homogeneous evolution (CHE) in both the core hydrogen and helium burning phase, 
as well as synthetic core helium burning stars with diverse hydrogen profiles which represent stripped stars.
In particular, we have determined the average slope of the hydrogen abundance profile in the envelopes
of the considered WR stars, which allowed us to identify the most likely evolutionary scenario
for producing the WR stars individually.

We found that in particular the two WR components of SMC\,AB5 (HD\,5980), but also the apparently single
WN star SMC\,AB12, have properties which are consistent with CHE. SMC\,AB5 is in fact difficult to explain in
any other way \citep{Koenigsberger14}. 

For the hot WR stars ($T_{\rm eff} > 60\,$kK), which are in the stage of core helium burning, we find
a dichotomy in the slope of the hydrogen profiles, with one group showing shallow slopes, which are
consistent with those found in stellar models at core hydrogen exhaustion, and the second group showing
much steeper slopes. A physical interpretation of this dichotomy, despite the small number of stars,
appears reasonable because the WR stars of the first group have massive O star companions, while no companions
have been found for any of the WR stars of the second group.

The WR stars with O star companions turn out to fit very well to models which undergo stable Roche lobe overflow.
This concerns the binary properties, in particular the mass ratio and the orbital period.
But it also holds for the properties of the WR stars, for which corresponding models do predict shallow
hydrogen profile slopes with values very close to the ones we derived.
 
When studying the evolution of the hydrogen profile slope in post-main-sequence stellar evolution models,
we found two possible ways for steep hydrogen profiles to develop. One is due to convective and
semiconvective mixing in the hydrogen-rich envelope, which occurs during the post-main-sequence
expansion of the star. The second is due to mass accretion during the main sequence stage, 
which leads to an increase of the convective core mass. We show that in the binary evolution context,
both possibilities point to the formation of the WR star in a common envelope phase, with a main sequence
star as the most likely companion in the first case, and --- excitingly --- a compact object in the second case.

A single star origin for the hot and apparently single SMC WR stars can not be firmly excluded. However, the high fraction of massive stars in close binaries (Sana et al. 2012), and the detailed properties of these WR stars render a common envelope evolution as the likely agent for removing their hydrogen-rich envelope (cf. Sect. 5.3). This would raise the WR binary fraction in the SMC from about 50\% \cite{Foellmi03} to near 100\%, in line with the expectation for a very metal poor environment.



A previous common envelope phase of the hot apparently single WR stars would imply that they presently do have companions. The prospects of finding these companions are in fact good. Due to their relatively weak winds, absorption lines are present in the spectra of many of them \citep[]{Marchenko07, Hainich15}, which may allow to push the current accuracy of the radial velocity measurements of $\sim$30\,\kms to much smaller values, such that even quite low mass companions might become detectable.

Finding these companions would in fact be very valuable, as they would give us the
very first observational benchmark for the efficiency of common envelope ejection in the
stellar mass range of black hole progenitors. Some of the companions may even be black holes ---
this cannot be excluded from the current upper limits of their X-ray emission \citep{Guerrero08a, Guerrero08b} ---
which would render such binaries as direct progenitors of massive double black hole systems.





\begin{acknowledgements}
The authors would like to thank Pablo Marchant for help and input and Nathan Grin and Thomas Tauris for fruitful discussions.
\end{acknowledgements}

\bibliographystyle{aa} 
\setlength{\bibsep}{0pt}
\bibliography{bib.bib}{}

\clearpage

\section*{Appendix A - Hertzsprung-Russell diagrams with typical inferred hydrogen slopes}
\renewcommand{\thefigure}{A\arabic{figure}}
\setcounter{figure}{0}


   \begin{figure}[ht]
   \centering
   \includegraphics[width = \linewidth]{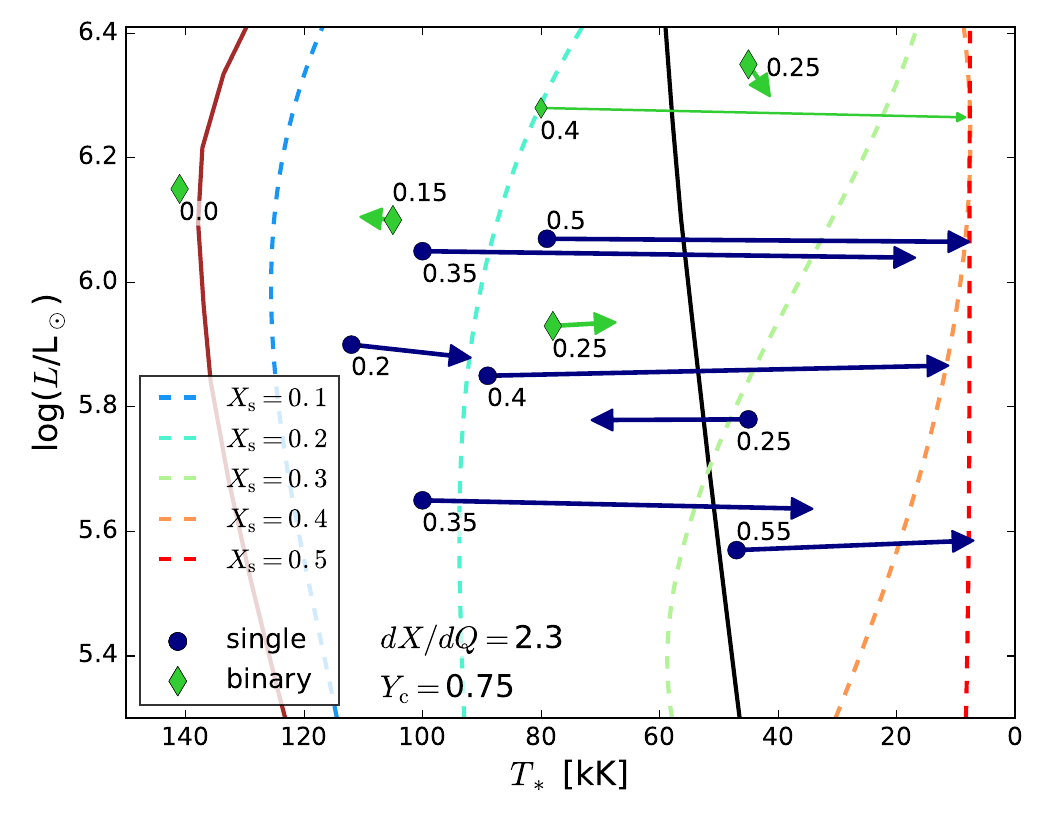}
   \caption{Hertzsprung-Russell diagram of models with helium cores and various hydrogen profiles displayed. The slope of the hydrogen profile has a value $dX/dQ = 2.3$ (which is the typical value we infer for the SMC WR stars in binaries) for all models. The surface hydrogen mass fraction $X_\mathrm{s}$ is indicated by the colors of the dashed lines.
   Green diamonds indicate observed values for binary SMC WR stars, blue circles indicate apparently singles stars.
   The arrows point to models with the same $\log L$ and $X_\mathrm{s}$ as the observed objects. The peculiar SMC\,AB6 is displayed with a smaller symbol and arrow.}
              \label{fig:stripped_star_hrd}%
    \end{figure}


   \begin{figure}[ht]
   \centering
   \includegraphics[width = \linewidth]{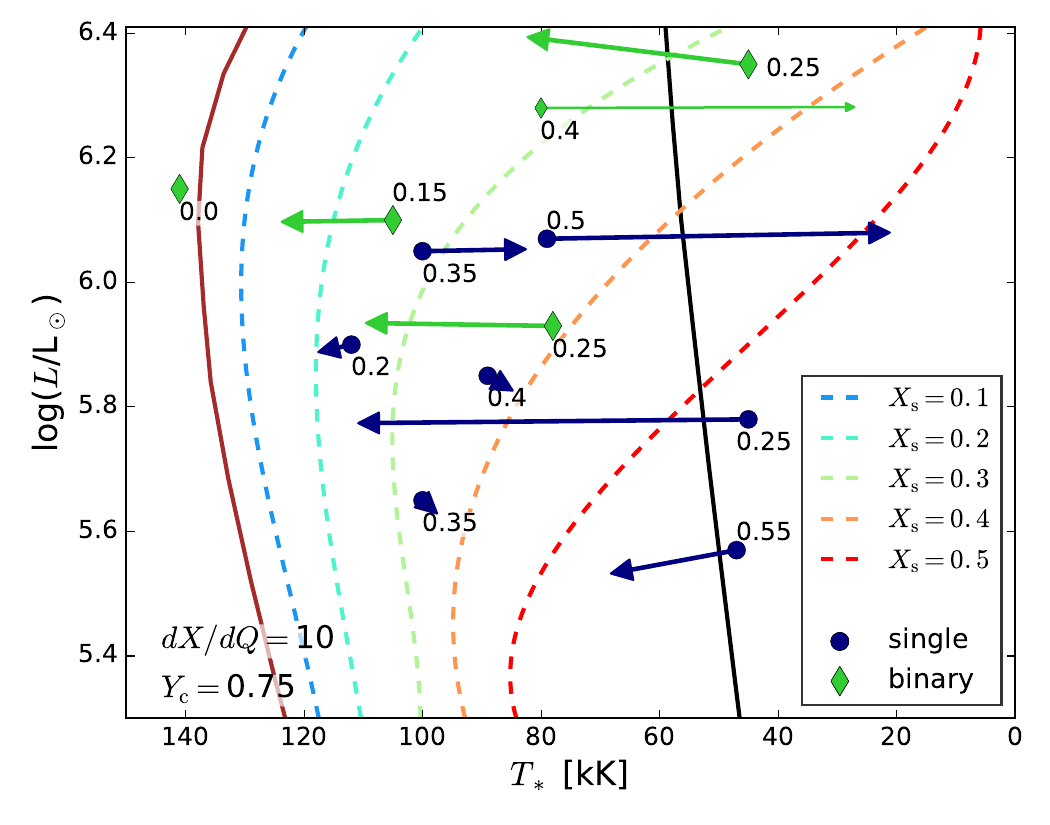}
   \caption{Same as Fig.\,\ref{fig:stripped_star_hrd}, but instead models with $dX/dQ = 10$ (which is the typical value we infer for the hot apparently single SMC WR stars) are shown.}
              \label{fig:stripped_star_hrd_noov}%
    \end{figure}

\clearpage

\section*{Appendix B - Best-fitting slopes for all WR stars\label{app:b}}

\renewcommand{\thefigure}{B\arabic{figure}}
\setcounter{figure}{0}


   \begin{figure}[ht]
   \centering
   \includegraphics[width =\linewidth]{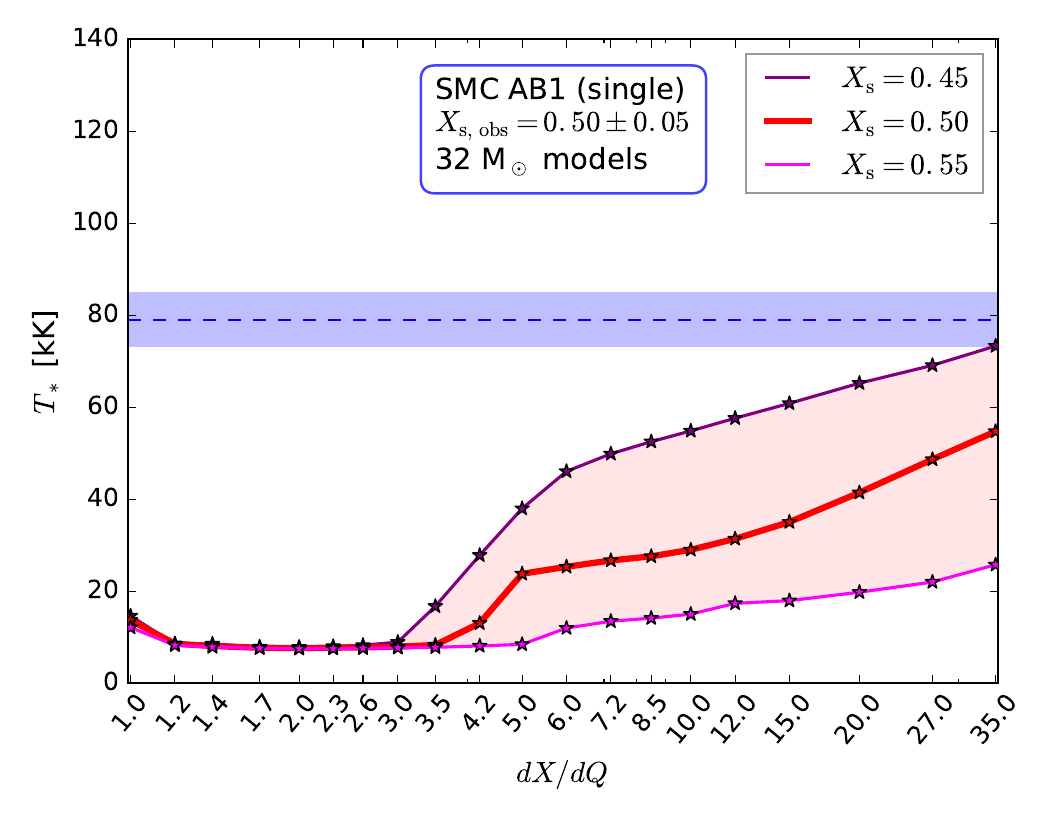}
   \caption{Same as Fig.\,\ref{fig:dxdq_teff11}, but models are compared to SMC\,AB1. \label{fig:dxdq_teff1}}%
    \end{figure}


   \begin{figure}[ht]
   \centering
   \includegraphics[width =\linewidth]{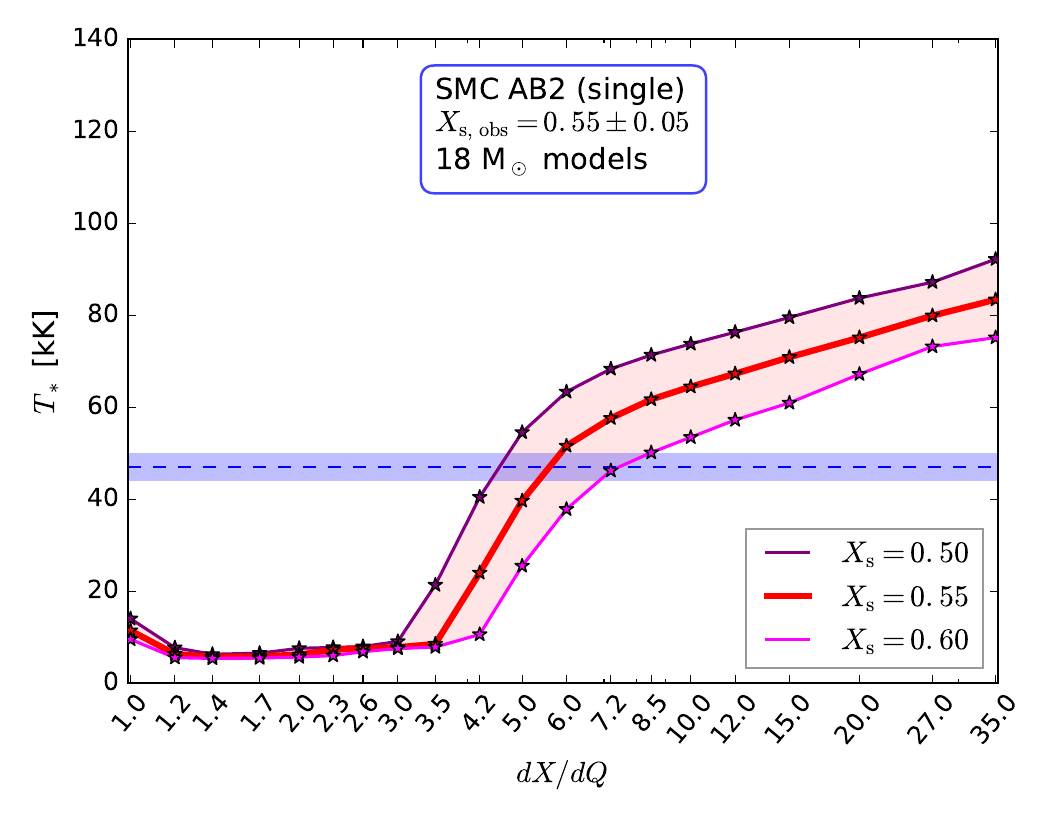}
   \caption{Same as Fig.\,\ref{fig:dxdq_teff11}, but models are compared to SMC\,AB2. \label{fig:dxdq_teff2} }%
    \end{figure}


   \begin{figure}[ht]
   \centering
   \includegraphics[width = \linewidth]{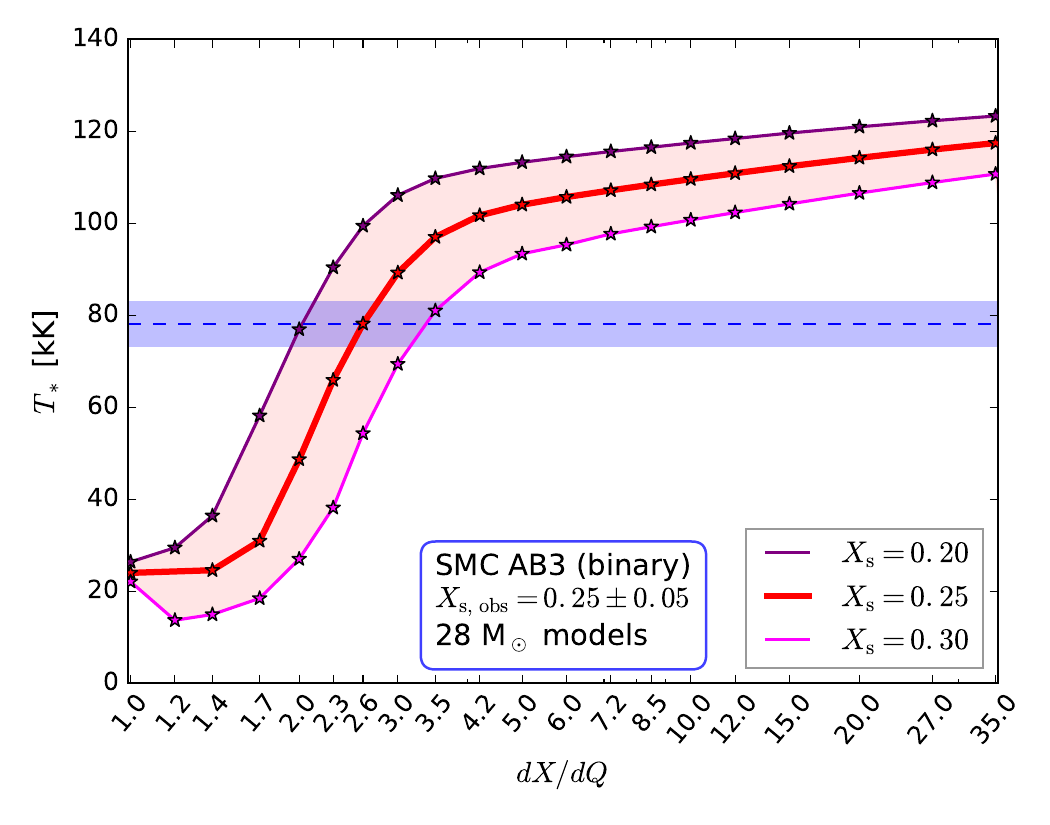}
   \caption{Same as Fig.\,\ref{fig:dxdq_teff11}, but models are compared to SMC\,AB3. \label{fig:dxdq_teff3}}%
    \end{figure}


   \begin{figure}[ht]
   \centering
   \includegraphics[width = \linewidth]{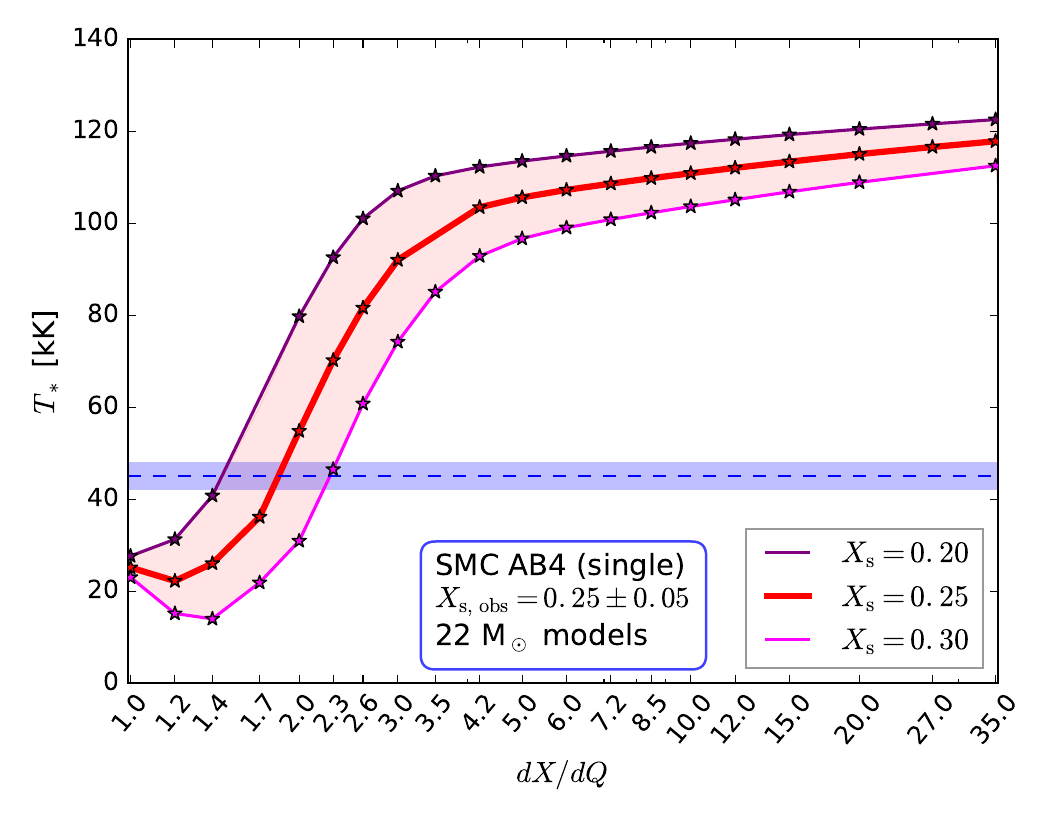}
   \caption{Same as Fig.\,\ref{fig:dxdq_teff11}, but models are compared to SMC\,AB4. \label{fig:dxdq_teff4}}%
    \end{figure}


   \begin{figure}[ht]
   \centering
   \includegraphics[width = 0.97\linewidth]{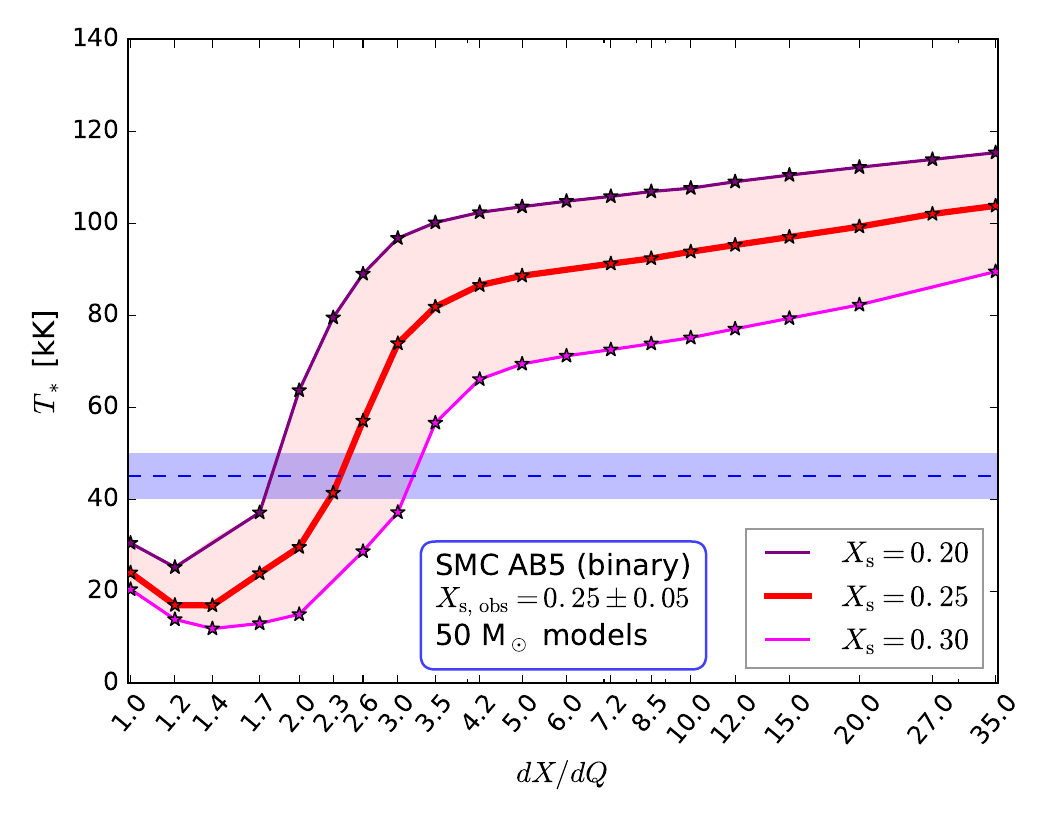}
   \caption{Same as Fig.\,\ref{fig:dxdq_teff11}, but models are compared to SMC\,AB5. \label{fig:dxdq_teff5}}%
    \end{figure}


   \begin{figure}[ht]
   \centering
   \includegraphics[width = 0.97\linewidth]{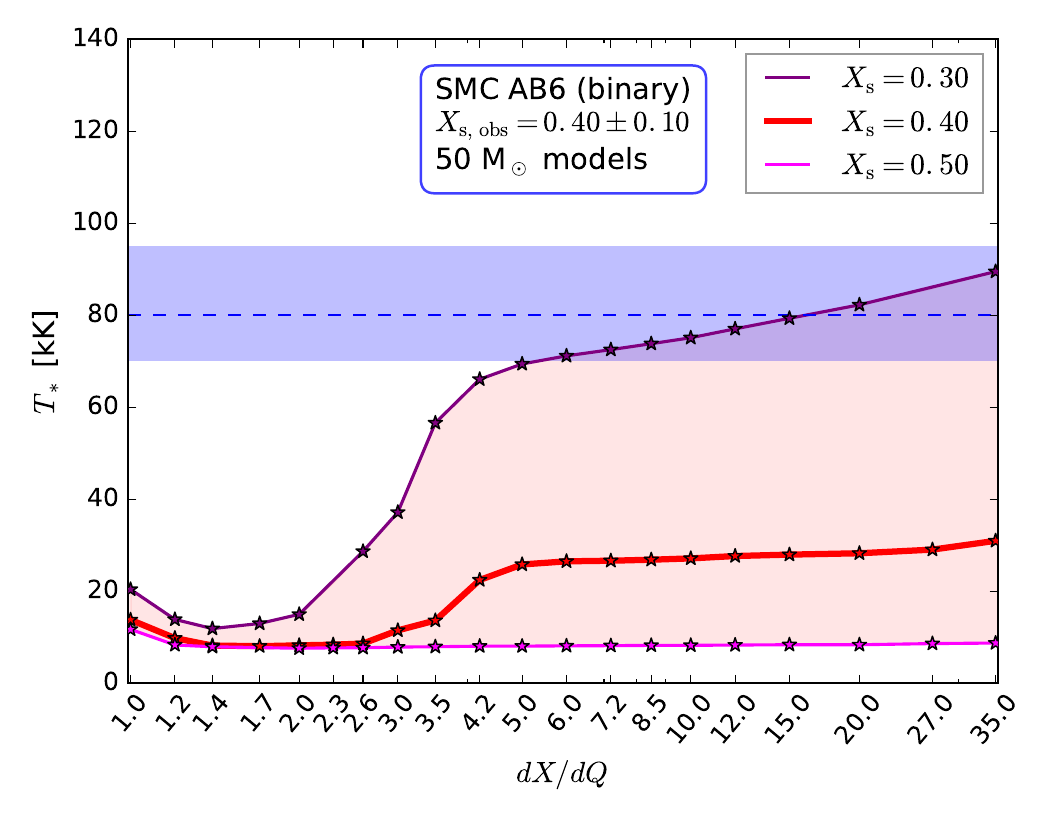}
   \caption{Same as Fig.\,\ref{fig:dxdq_teff11}, but models are compared to SMC\,AB6. \label{fig:dxdq_teff6}}%
    \end{figure}


   \begin{figure}
   \centering
   \includegraphics[width = 0.97\linewidth]{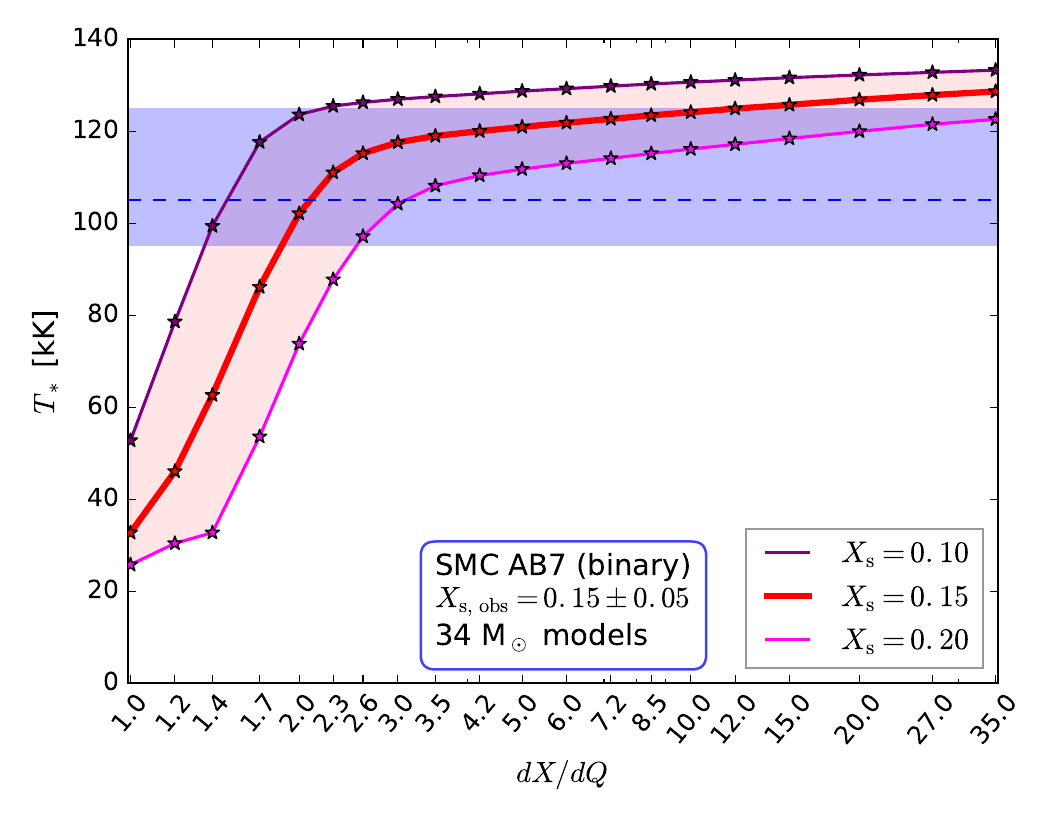}
   \caption{Same as Fig.\,\ref{fig:dxdq_teff11}, but models are compared to SMC\,AB7. \label{fig:dxdq_teff7}}%
    \end{figure}


   \begin{figure}[ht]
   \centering
   \includegraphics[width = 0.97\linewidth]{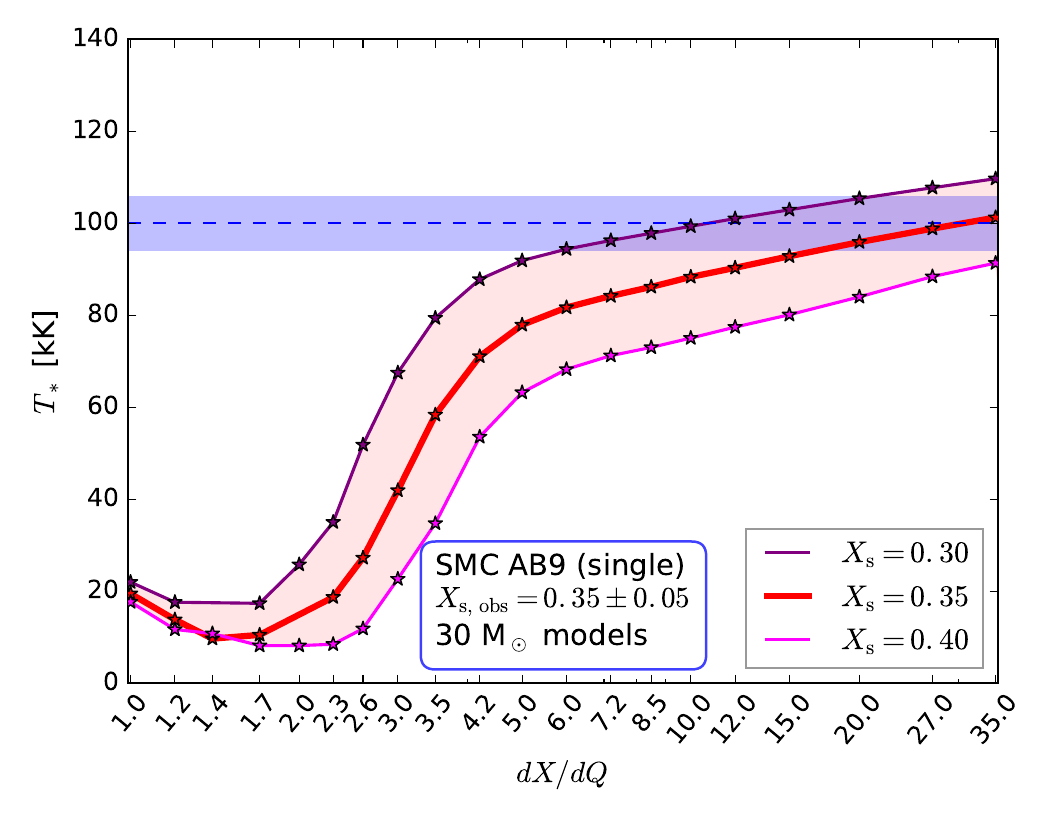}
   \caption{Same as Fig.\,\ref{fig:dxdq_teff11}, but models are compared to SMC\,AB9. \label{fig:dxdq_teff9}}%
    \end{figure}


   \begin{figure}[ht]
   \centering
   \includegraphics[width = 0.97\linewidth]{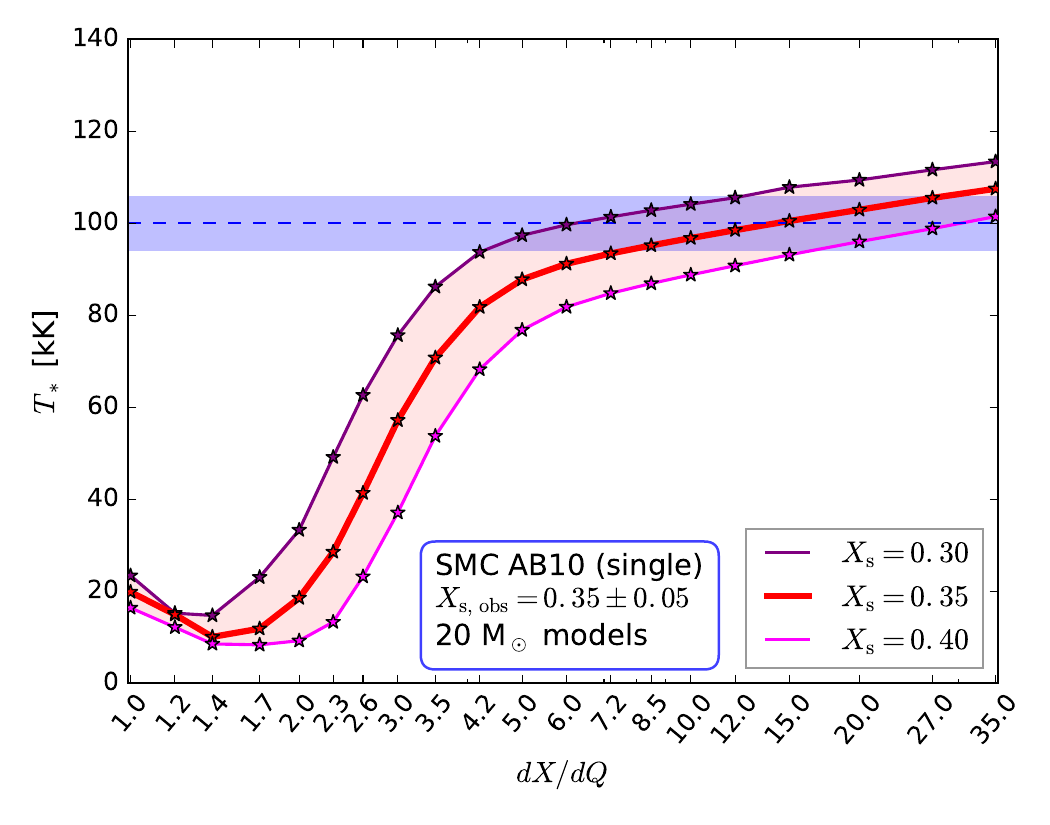}
   \caption{Same as Fig.\,\ref{fig:dxdq_teff11}, but models are compared to SMC\,AB10.) \label{fig:dxdq_teff10}}%
    \end{figure}


   \begin{figure}[ht]
   \centering
   \includegraphics[width = 0.97\linewidth]{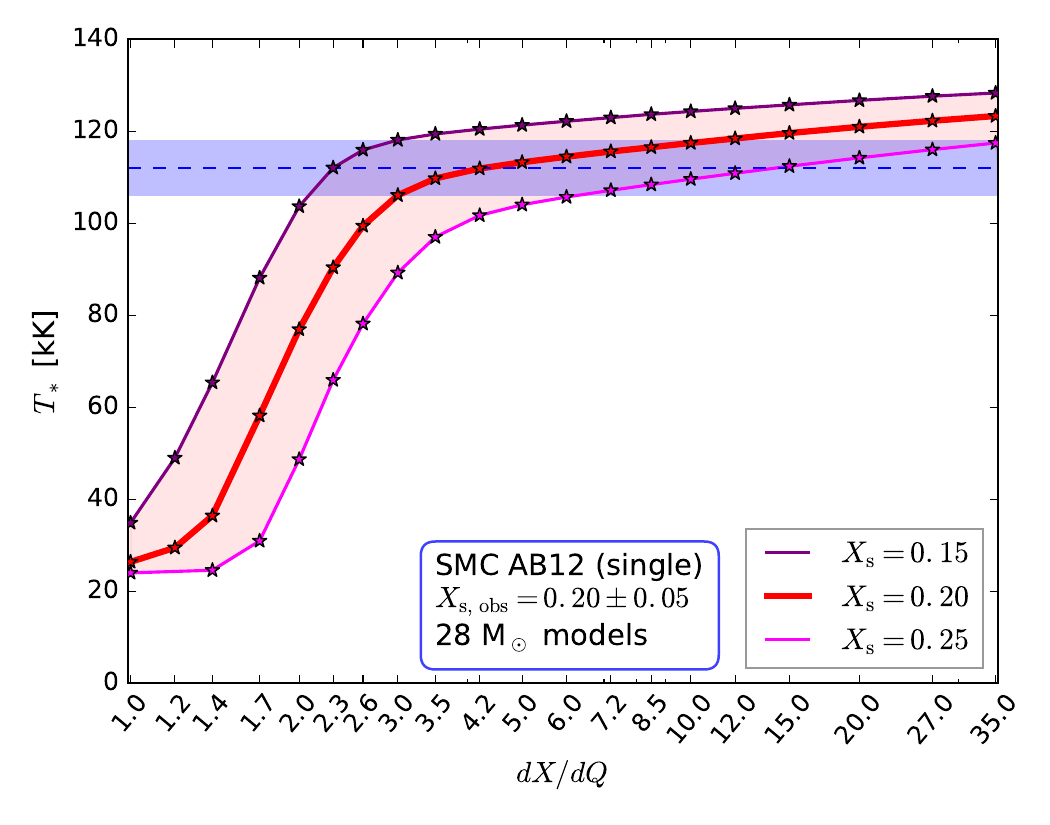}
   \caption{Same as Fig.\,\ref{fig:dxdq_teff11}, but models are compared to SMC\,AB12.\label{fig:dxdq_teff12}}%
    \end{figure}

\clearpage
\section*{Appendix C - Validity of the method \label{app:c}}

\renewcommand{\thefigure}{C\arabic{figure}}
\setcounter{figure}{0}


   \begin{figure}[ht]
   \centering
   \includegraphics[width =\linewidth]{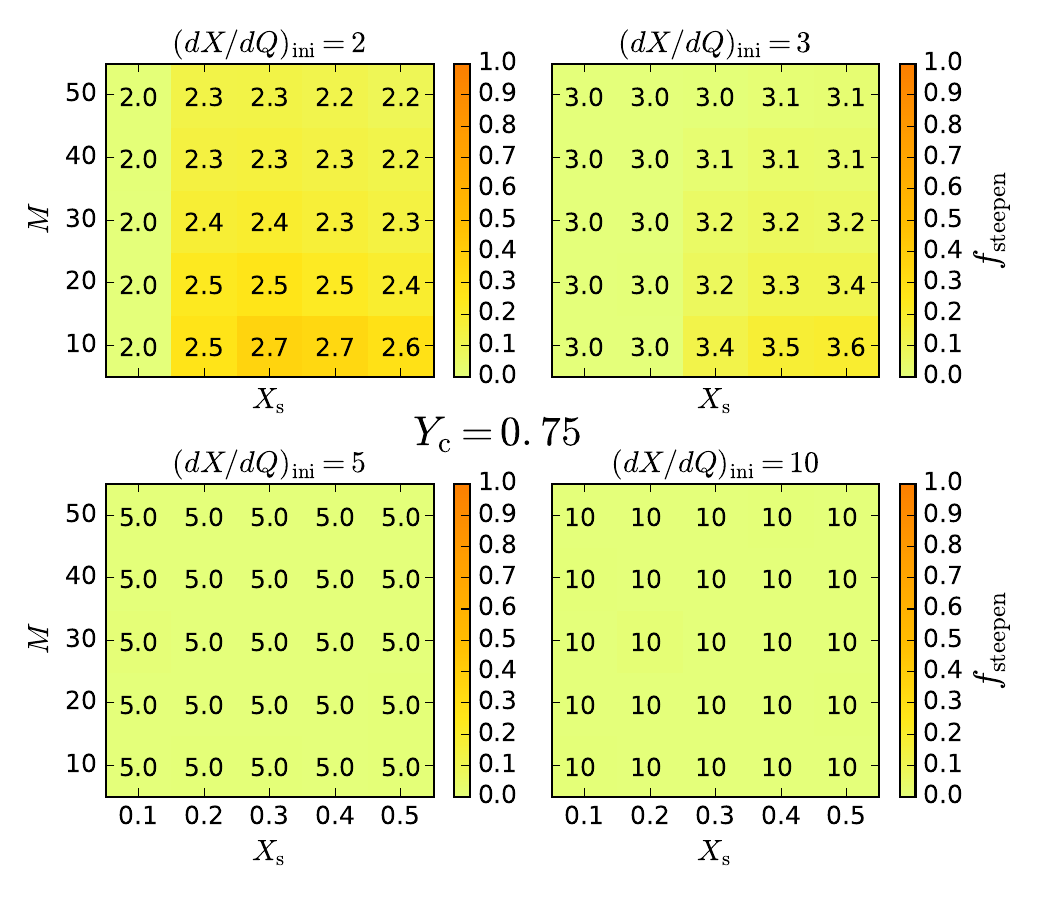}
   \caption{Diagrams showing the change of the hydrogen slope $dX/dQ$ as a result of hydrogen shell burning. Each number indicates the values of $dX/dQ$ at the moment helium burning has proceeded to $Y_\mathrm{c} = 0.75$ in models with different initial values for $dX/dQ$, $X_\mathrm{s}$ and $M$. Unlike the models discussed in Sect.\,\ref{sec:other_channels}, hydrogen shells burning is allowed in these models. The background color indicates by what fraction $dX/dQ$ has increased during helium burning: $f_\mathrm{steepen}$. \label{fig:evol_dxdqs_final}}%
    \end{figure}

\noindent 
In this appendix, we consider the effect of hydrogen shell burning on the $dX/dQ$ value of the hydrogen profiles in our stripped star models and the timescales on which the hydrogen envelopes we infer would be blown away by wind mass loss.
We find that although hydrogen shell burning occurs, it does not dramatically increase the $dX/dQ$ value during helium burning. This is shown in Fig.\,\ref{fig:evol_dxdqs_final} in the appendix: typically the $dX/dQ$ value increases by less than 25$\%$ until $Y_\mathrm{c} = 0.75$. The largest increase that occurs in the whole parameter space is from $dX/dQ = 2.0$ to $dX/dQ = 4.1$ over the entire core helium burning phase. Models with higher initial $dXdQ$ values are less affected by nuclear burning.

In an extreme case with a strong stellar wind and a low-mass hydrogen envelope (the higher $dX/dQ$ and the lower $X_\mathrm{s}$, the lower the mass of the hydrogen envelope), it is imaginable that the hydrogen envelope is completely removed during core helium burning. Therefore we consider the lifetimes of the model hydrogen envelopes, $\tau_\mathrm{H \, env} = M_\mathrm{H \, env}/ \dot{M}_\mathrm{obs}$, with respect to the model core helium burning lifetimes, $\tau_\mathrm{He \, core}$. Here, $M_\mathrm{H \, env}$ follows from the inferred $dX/dQ$ value, the observed hydrogen mass fraction $X_\mathrm{s}$ and the mass $M_\mathrm{model}$ that corresponds to the observed luminosity. In a case where $\tau_\mathrm{H \, env} / \tau_\mathrm{He \, core} > 1$, there is no moment during helium burning in which the star is hydrogen free. What we find is that the ratio $\tau_\mathrm{H \, env} / \tau_\mathrm{He \, core}$ is typically of the order unity (Table\,\ref{tab:dxdq_fit}). That means that for some stars where the value is below one, the hydrogen-depleted layers would be exposed at a point in time, $-$ given that envelope stripping didn't happen in a late helium burning phase. We note that more efficient convection (a higher value of $\alpha_\mathrm{MLT}$) would lead to less envelope inflation and somewhat higher values for $T_\mathrm{*}$ mainly in the case of luminous and hydrogen-rich stars (i.e., SMC\,AB1, 6 and 9). A quick test showed that this results in slightly lower $dX/dQ$ values also being compatible with the observed properties, and allows for higher $\tau_\mathrm{H \, env}$. However, this did not change the conclusion that these objects are incompatible with models where $dX/dQ$ is on the order of 2: their $T_*$ was not significantly affected.


   \begin{figure}
   \centering
   \includegraphics[width =\linewidth]{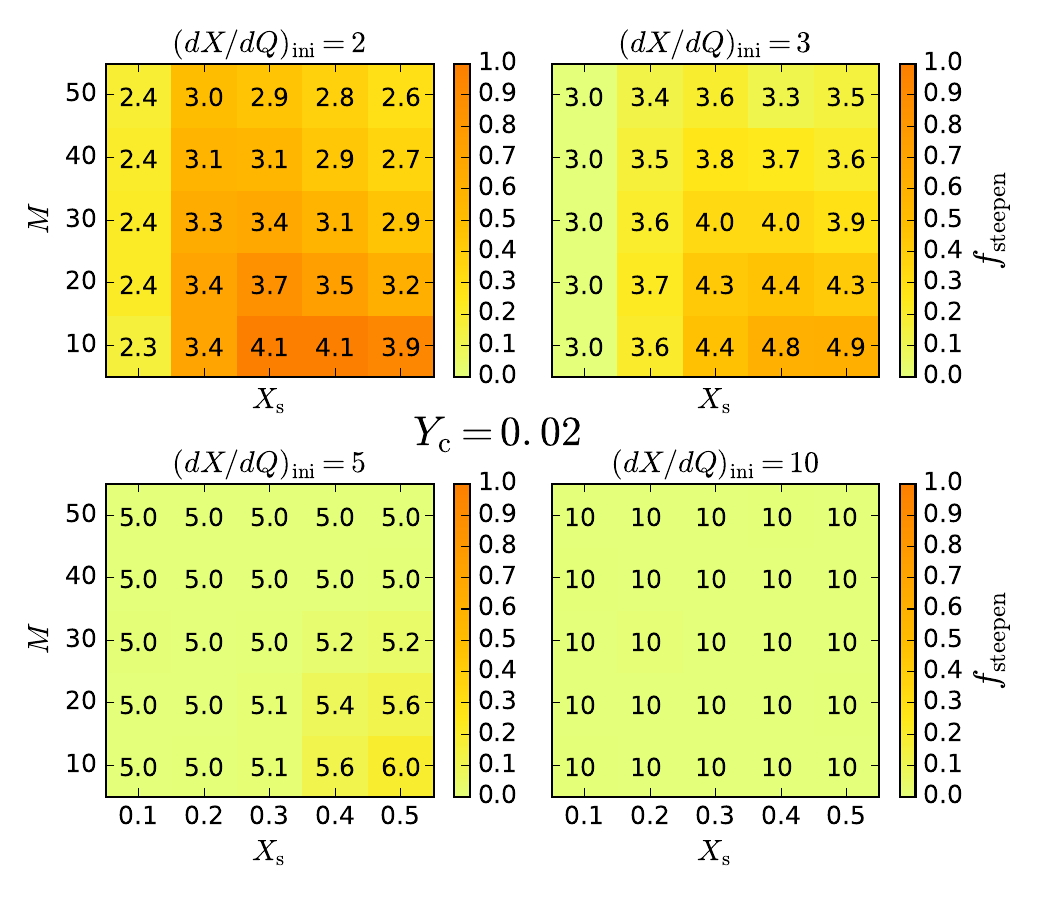}
   \caption{Same as Fig.\,\ref{fig:evol_dxdqs_final}, but now the models are almost at the end of helium burning: $Y_\mathrm{c} = 0.02$.}%
    \end{figure}
\renewcommand{\thetable}{C\arabic{table}}
\setcounter{table}{0}
\begin{table}
\centering    
    \caption[1.4\linewidth]{Best-fitting $dX/dQ$ values inferred for the single and binary SMC WR stars. Also displayed are the timescale on which the WR star would blow away its hydrogen-containing layers ($\tau_\mathrm{H \, env}$) and the ratio of $\tau_\mathrm{H \, env}$ to the helium burning timescale. The last column gives the best fitting
current masses.
    }
\begin{tabular}{l l l l l}
\hline
\hline
            \noalign{\smallskip}
            SMC & $dX/dQ$ & $\tau_\mathrm{H \, env}$ & \multirow{2}{*}{ $\frac{
            \tau_\mathrm{H \, env} }{
            \tau_\mathrm{He \, burn}
            }$
            } & $M_\mathrm{model}$
            \\
            AB & & [kyr] &  & [\Msolend]\\
            \noalign{\smallskip}
            \hline
            \noalign{\smallskip}
            \multicolumn{2}{l}{\textbf{Single:}}\\
            \noalign{\smallskip}
            1 & $\geq 35$ & $ \leq 175$ & $\leq 0.56$ & 32\\
            \noalign{\smallskip}
            2 & $5.7^{+1.8}_{-1.2}$ & $1011^{+282}_{-299}$ & $2.66^{+0.74}_{-0.65}$ & 18\\
            \noalign{\smallskip}
            4 & $1.8^{+0.5}_{-0.3}$ & $638^{+191}_{-177}$ & $1.81^{+0.55}_{-0.50}$ & 22\\
            \noalign{\smallskip}
            9 & $30^{+?}_{-18.5}$ & $158^{+264}_{-?}$ & $0.50^{+0.81}_{-?}$ & 30\\
            \noalign{\smallskip}
            10 & $13.5^{+?}_{-7.5}$ & $233^{+312}_{-?}$ & $0.64^{+0.85}_{-?}$ & 20\\
            \noalign{\smallskip}
            11 & $18^{+17}_{-11}$ & $213^{+354}_{-86}$ & $0.64^{+1.06}_{-0.26}$ & 26\\
            \noalign{\smallskip}
            12 & $4.2^{+10.8}_{-1.9}$ & $945^{+1023}_{-707}$ & $2.91^{+3.15}_{-2.18}$ & 28\\
            \noalign{\smallskip}
            \hline
            \noalign{\smallskip}
            \multicolumn{2}{l}{\textbf{Binary:}}\\
            \noalign{\smallskip}
            3 & $2.6^{+0.7}_{-0.5}$ & $663^{+207}_{-165}$ & $2.04^{+0.64}_{-1.53}$ & 28\\
            \noalign{\smallskip}
            5 & $2.4^{+0.8}_{-0.6}$ & $207^{+96}_{-61}$ & $0.75^{+0.34}_{-0.22}$ & 50\\
            \noalign{\smallskip}
            6 & $\geq 16$ & $ \leq 160$ & $\leq 0.58$ & 50\\
            \noalign{\smallskip}
            7 & $2.1^{+0.9}_{-0.6}$ & $188^{+168}_{-55}$ & $0.
            61^{+0.53}_{-0.18}$ & 34\\
            \noalign{\smallskip}

            \hline
\end{tabular}
\label{tab:dxdq_fit}
\end{table}

\clearpage

\renewcommand{\thetable}{D\arabic{table}}
\setcounter{table}{0}

\section*{Appendix D \label{app:d}}
\noindent\begin{minipage}{\textwidth}
\centering    
    \captionof{table}{
    Models which are able to achieve the best fit for each individual Wolf-Rayet star. For each object we display the initial mass and initial rotation velocity of the best-fitting models and we compare the observed values \citep{Hainich15} of the fit parameters to their model values at the moment the best fit is achieved. Next, we display the observed upper limit on the rotational velocity $v \sin i$ and the model $v_\mathrm{rot}$ at the moment the best fit is achieved. For models where $v_\mathrm{rot}$ exceeds the observed upper limit on $v \sin i$ we calculate $P_\mathrm{inc}$, i.e., the chance that $v \sin i$ does not exceed the upper limit due to a low inclination of the rotational axis. $\chi^2_\mathrm{min}$ is the lowest $\chi^2$ value achieved with our three fit parameters $T_*$, $\log L$ and $X_\mathrm{H}$. In this table, only \textbf{single stars} are considered and compared to models that are \textbf{core hydrogen burning}.
    }
\begin{tabular}{l l l l l l l l l}
\hline
\hline
            \noalign{\smallskip}
            SMC & $M_0$ & $v_\mathrm{rot, 0}$ & $T_*$ & $\log L$ & $X_\mathrm{H}$ & $v \sin i$; $v_\mathrm{rot}$ & $P_\mathrm{inc}$ & $\chi^2_\mathrm{min}$\\
            AB & [\Msol ] & [km s$^{-1}$] & [kK] & [\Lsol] & & [km s$^{-1}$] & &\\
            \noalign{\smallskip}
            \hline
            \noalign{\smallskip}
            
            \textbf{1} (obs) & & & $79^{+6}_{-6}$ &$6.07^{+0.20}_{-0.20}$ & $0.50^{+0.05}_{-0.05}$  & $<$100 & \multirow{2}{*}{0.044} & \multirow{2}{*}{11.6}\\
            {\tt model} & {\tt 100} & {\tt 590} & {\tt 59} & {\tt 6.25} & {\tt 0.48}  & {\tt 342} & & \\
            \noalign{\smallskip}
            \noalign{\smallskip}           
            
            \textbf{2} (obs) & & & $47^{+3}_{-3}$ &$5.57^{+0.10}_{-0.10}$ & $0.55^{+0.05}_{-0.05}$  & $<$50 & \multirow{2}{*}{0.014} & \multirow{2}{*}{0.29} \\
            {\tt model} & {\tt 35} & {\tt 400} & {\tt 47} & {\tt 5.54} & {\tt 0.53}  & {\tt 302} & & \\
            \noalign{\smallskip}
            \noalign{\smallskip}
            
            \textbf{4} (obs) & & & $45^{+3}_{-3}$ &$5.78^{+0.10}_{-0.10}$ & $0.25^{+0.05}_{-0.05}$  & $<$100 & \multirow{2}{*}{0.16} & \multirow{2}{*}{3.07}\\
            {\tt model} & {\tt 35} & {\tt 420} & {\tt 50} & {\tt 5.81} & {\tt 0.29}  & {\tt 183} & & \\
            \noalign{\smallskip}
            \noalign{\smallskip}            
            
            \textbf{9} (obs) & & & $100^{+6}_{-6}$ &$6.05^{+0.20}_{-0.20}$ & $0.35^{+0.05}_{-0.05}$  & $<$200 & \multirow{2}{*}{0.40} & \multirow{2}{*}{45.3} \\
            {\tt model} & {\tt 70} & {\tt 600} & {\tt 60} & {\tt 6.12} & {\tt 0.32}  & {\tt 251} & & \\
            \noalign{\smallskip}
            \noalign{\smallskip}
            
            \textbf{10} (obs) & & & $100^{+6}_{-6}$ &$5.65^{+0.20}_{-0.20}$ & $0.35^{+0.05}_{-0.05}$  & $<$200 & \multirow{2}{*}{0.36} & \multirow{2}{*}{49.6} \\
            {\tt model} & {\tt 55} & {\tt 600} & {\tt 60} & {\tt 5.98} & {\tt 0.29}  & {\tt 259} & & \\
            \noalign{\smallskip}
            \noalign{\smallskip}
            
            \textbf{11} (obs) & & & $89^{+6}_{-6}$ &$5.85^{+0.20}_{-0.20}$ & $0.40^{+0.05}_{-0.05}$  & $<$200 & \multirow{2}{*}{0.34} & \multirow{2}{*}{26.1} \\
            {\tt model} & {\tt 100} & {\tt 600} & {\tt 62} & {\tt 6.31} & {\tt 0.37}  & {\tt 267} & & \\
            \noalign{\smallskip}
            \noalign{\smallskip}
            
            \textbf{12} (obs) & & & $112^{+6}_{-6}$ &$5.90^{+0.20}_{-0.20}$ & $0.20^{+0.05}_{-0.05}$  & $<$200 & \multirow{2}{*}{1} & \multirow{2}{*}{49.9} \\
            {\tt model} & {\tt 100} & {\tt 600} & {\tt 77} & {\tt 6.31} & {\tt 0.02}  & {\tt 13} & & \\
            \noalign{\smallskip}
            \noalign{\smallskip}  
            
            \hline
\end{tabular}
\label{tab:chi2_ss_h}


\vspace{5mm}

\centering    
    \captionof{table}{Lowest obtained $\chi^2$ values for each system.
    Same as Table\,\ref{tab:chi2_ss_h}, but now we compare with \textbf{core helium burning} models.
    }
\begin{tabular}{l l l l l l l l l}
\hline
\hline
            \noalign{\smallskip}
            SMC & $M_0$ & $v_\mathrm{rot, 0}$ & $T_*$ & $\log L$ & $X_\mathrm{H}$ & $v \sin i$; $v_\mathrm{rot}$ & $P_\mathrm{inc}$ & $\chi^2_\mathrm{min}$\\
            AB & [\Msol ] & [km s$^{-1}$] & [kK] & [\Lsol] & & [km s$^{-1}$] & &\\
            \noalign{\smallskip}
            \hline
            \noalign{\smallskip}
            
            \textbf{1} (obs) & & & $79^{+6}_{-6}$ &$6.07^{+0.20}_{-0.20}$ & $0.50^{+0.05}_{-0.05}$  & $<$100 & \multirow{2}{*}{1} & \multirow{2}{*}{8.5}\\
            {\tt model} & {\tt 55} & {\tt 390} & {\tt 67} & {\tt 6.15} & {\tt 0.40}  & {\tt 0.2} & & \\
            \noalign{\smallskip}
            \noalign{\smallskip}           
            
            \textbf{2} (obs) & & & $47^{+3}_{-3}$ &$5.57^{+0.10}_{-0.10}$ & $0.55^{+0.05}_{-0.05}$  & $<$50 & \multirow{2}{*}{1} & \multirow{2}{*}{39.3} \\
            {\tt model} & {\tt 55} & {\tt 390} & {\tt 39} & {\tt 6.15} & {\tt 0.44}  & {\tt 1} & & \\
            \noalign{\smallskip}
            \noalign{\smallskip}
            
            \textbf{4} (obs) & & & $45^{+3}_{-3}$ &$5.78^{+0.10}_{-0.10}$ & $0.25^{+0.05}_{-0.05}$  & $<$100 & \multirow{2}{*}{1} & \multirow{2}{*}{1.8}\\
            {\tt model} & {\tt 30} & {\tt 450} & {\tt 45} & {\tt 5.9} & {\tt 0.22}  & {\tt 79} & & \\
            \noalign{\smallskip}
            \noalign{\smallskip}            
            
            \textbf{9} (obs) & & & $100^{+6}_{-6}$ &$6.05^{+0.20}_{-0.20}$ & $0.35^{+0.05}_{-0.05}$  & $<$200 & \multirow{2}{*}{1} & \multirow{2}{*}{8.4} \\
            {\tt model} & {\tt 45} & {\tt 550} & {\tt 100} & {\tt 6.08} & {\tt 0.21}  & {\tt 6} & & \\
            \noalign{\smallskip}
            \noalign{\smallskip}      
            
            \textbf{10} (obs) & & & $100^{+6}_{-6}$ &$5.65^{+0.20}_{-0.20}$ & $0.35^{+0.05}_{-0.05}$  & $<$200 & \multirow{2}{*}{1} & \multirow{2}{*}{11.3} \\
            {\tt model} & {\tt 20} & {\tt 460} & {\tt 99} & {\tt 5.65} & {\tt 0.18}  & {\tt 5} & & \\
            \noalign{\smallskip}
            \noalign{\smallskip}
            
            \textbf{11} (obs) & & & $89^{+6}_{-6}$ &$5.85^{+0.20}_{-0.20}$ & $0.40^{+0.05}_{-0.05}$  & $<$200 & \multirow{2}{*}{1} & \multirow{2}{*}{9.3} \\
            {\tt model} & {\tt 55} & {\tt 390} & {\tt 75} & {\tt 6.15} & {\tt 0.34}  & {\tt 9} & & \\
            \noalign{\smallskip}
            \noalign{\smallskip}
            
            \textbf{12} (obs) & & & $112^{+6}_{-6}$ &$5.90^{+0.20}_{-0.20}$ & $0.20^{+0.05}_{-0.05}$  & $<$200 & \multirow{2}{*}{1} & \multirow{2}{*}{0.7} \\
            {\tt model} & {\tt 30} & {\tt 510} & {\tt 111} & {\tt 5.88} & {\tt 0.16}  & {\tt 8} & & \\
            \noalign{\smallskip}
            \noalign{\smallskip}              
            \hline
\end{tabular}
\label{tab:chi2_ss_he}
\end{minipage}

\clearpage

\begin{table*}[ht]
\centering    
    \caption[1.4\linewidth]{
    Models which are able to achieve the best fit for each individual Wolf-Rayet star. For each object we display the initial mass and initial rotation velocity of the best-fitting models and we compare the observed values \citep{Shenar16} of the fit parameters to their model values at the moment the best fit is achieved. The parameter $v_\mathrm{sync}$ is the rotational velocity that the star would have in a system with (tidally) synchronized orbital and rotation periods. $R_\mathrm{RL}$ is the current size of the star's Roche lobe, whereas $R_\mathrm{max}$ is the maximum radius of the models at any point in time before the best fit is achieved. $\chi^2_\mathrm{min}$ is the lowest $\chi^2$ value achieved with our three fit parameters $T_*$, $\log L$ and $X_\mathrm{H}$. Unlike in Table\,\ref{tab:chi2_ss_h}, we do not consider the model rotation velocities since these objects do not have observed constraints on the rotation velocity (except SMC\,AB 5$_\mathrm{A}$ with $v \sin i < 300$\kms). In this table, only \textbf{binary stars} are considered and compared to models that are \textbf{core hydrogen burning}.
    }
\begin{tabular}{l l l l l l l l l l}
\hline
\hline
            \noalign{\smallskip}
            SMC & $M_0$ & $v_\mathrm{rot, 0}$ & $T_*$ & $\log L$ & $X_\mathrm{H}$ & $v_\mathrm{sync}$ & $R_\mathrm{RL}$ & $R_\mathrm{max}$ & $\chi^2_\mathrm{min}$\\
            AB & [\Msolend ] & [km s$^{-1}$] & [kK] & [\Lsolend] &  & [km s$^{-1}$]& [\Rsolend ] & [\Rsolend ] & \\
            \noalign{\smallskip}
            \hline
            \noalign{\smallskip}
            
            \textbf{3} (obs) & & & $78^{+5}_{-5}$ &$5.93^{+0.05}_{-0.05}$ & $0.25^{+0.05}_{-0.05}$ & $25^{+5}_{-5}$ & $25^{+29}_{-6}$ & & \multirow{2}{*}{13.3} \\
            {\tt model} & {\tt 50} & {\tt 600} & {\tt 60} & {\tt 5.96} & {\tt 0.25} &  & & {\tt 9.3} & \\
            \noalign{\smallskip}
            \noalign{\smallskip}           
            
            \textbf{$5_\mathrm{A}$} (obs) & & & $45^{+5}_{-5}$ &$6.35^{+0.10}_{-0.10}$ & $0.25^{+0.05}_{-0.05}$ & $63^{+26}_{-18}$ & $58^{+4}_{-4}$ & & \multirow{2}{*}{1.0} \\
            {\tt model} & {\tt 80} & {\tt 540} & {\tt 45} & {\tt 6.29} & {\tt 0.29} & &  & {\tt 24} & \\
            \noalign{\smallskip}
            \noalign{\smallskip}      
            
            \textbf{6} (obs) & & & $80^{+15}_{-10}$ &$6.28^{+0.10}_{-0.10}$ & $0.40^{+0.10}_{-0.10}$ & $54^{+24}_{-15}$ & $14^{+3}_{-2}$ & &  \multirow{2}{*}{6.75} \\
            {\tt model} & {\tt 100} & {\tt 600} & {\tt 63} & {\tt 6.33} & {\tt 0.34} & & & {\tt 13.1} & \\
            \noalign{\smallskip}
            \noalign{\smallskip}
            
            \textbf{7} (obs) & & & $105^{+20}_{-10}$ &$6.10^{+0.10}_{-0.10}$ & $0.15^{+0.05}_{-0.05}$ & $9^{+3}_{-3}$ & $40^{+8}_{-3}$ & & \multirow{2}{*}{18.4} \\
            \noalign{\smallskip}
            {\tt model} & {\tt 100} & {\tt 600} & {\tt 77} & {\tt 6.31} & {\tt 0.02} & & & {\tt 13.1} & \\
            \noalign{\smallskip}
            \noalign{\smallskip}

            \hline
\end{tabular}
\label{tab:chi2_bin_h}
\end{table*}


\begin{table*}[ht]
\centering    
    \caption[1.4\linewidth]{Same as Table\,\ref{tab:chi2_bin_h}, but now we compare with \textbf{core helium burning} models.
    }
\begin{tabular}{l l l l l l l l l l}
\hline
\hline
            \noalign{\smallskip}
            SMC & $M_0$ & $v_\mathrm{rot, 0}$ & $T_*$ & $\log L$ & $X_\mathrm{H}$ & $v_\mathrm{sync}$ & $R_\mathrm{RL}$ & $R_\mathrm{max}$ & $\chi^2_\mathrm{min}$\\
            AB & [\Msolend ] & [km s$^{-1}$] & [kK] & [\Lsolend] &  & [km s$^{-1}$]& [\Rsolend ] & [\Rsolend ] & \\
            \noalign{\smallskip}
            \hline
            \noalign{\smallskip}
            
            \textbf{3} (obs) & & & $78^{+5}_{-5}$ &$5.93^{+0.05}_{-0.05}$ & $0.25^{+0.05}_{-0.05}$ & $25^{+5}_{-5}$ & $25^{+29}_{-6}$ & & \multirow{2}{*}{0.5} \\
            {\tt model} & {\tt 35} & {\tt 480} & {\tt 78} & {\tt 5.96} & {\tt 0.23} & & & {\tt 27} & \\
            \noalign{\smallskip}
            \noalign{\smallskip}           
            
            \textbf{$5_\mathrm{A}$} (obs) & & & $45^{+5}_{-5}$ &$6.35^{+0.10}_{-0.10}$ & $0.25^{+0.05}_{-0.05}$ & $63^{+26}_{-18}$ & $58^{+4}_{-4}$ & & \multirow{2}{*}{0.04} \\
            {\tt model} & {\tt 70} & {\tt 520} & {\tt 45} & {\tt 6.33} & {\tt 0.25} & & & {\tt 103} & \\
            \noalign{\smallskip}
            \noalign{\smallskip}      
            
            \textbf{6} (obs) & & & $80^{+15}_{-10}$ &$6.28^{+0.10}_{-0.10}$ & $0.40^{+0.10}_{-0.10}$ & $54^{+24}_{-15}$ & $14^{+3}_{-2}$ & & \multirow{2}{*}{2.5} \\
            {\tt model} & {\tt 55} & {\tt 390} & {\tt 75} & {\tt 6.13} & {\tt 0.34} & & & {\tt 1550} & \\
            \noalign{\smallskip}
            \noalign{\smallskip}
            
            \textbf{7} (obs) & & & $105^{+20}_{-10}$ &$6.10^{+0.10}_{-0.10}$ & $0.15^{+0.05}_{-0.05}$ & $9^{+3}_{-3}$ & $40^{+8}_{-3}$ & & \multirow{2}{*}{0.03} \\
            \noalign{\smallskip}
            {\tt model} & {\tt 45} & {\tt 530} & {\tt 106} & {\tt 6.10} & {\tt 0.16} & & & {\tt 33} & \\
            \noalign{\smallskip}
            \noalign{\smallskip}

            \hline
\end{tabular}
\label{tab:chi2_bin_he}
\end{table*}

\end{document}